\documentclass[twocolumn]{aastex63}

\usepackage{xspace}

\newcommand{\todo}{\ifmmode \text{\color{purple}\Huge{\(\bullet\)}} \else {\color{purple}{\Huge$\bullet$}}\fi}
\newcommand{\finish}{\ifmmode \text{\color{blue}\Huge{\(\bullet\)}} \else {\color{blue}{\Huge$\bullet$}}\fi}

\newcommand{\R}{\mathcal{R_{\rm rest}}}

\definecolor{darkgoldenrod}{rgb}{0.72, 0.53, 0.04}

\usepackage{natbib}
\usepackage{amsmath}
\usepackage{multirow} %for using \multirow
\usepackage{ifthen} %if/then package
\usepackage{txfonts}
\usepackage{tabularx}
\received{\today}
\usepackage{lineno}
\submitjournal{ApJS}

\shorttitle{HSC-VLASS dropout catalog}
\shortauthors{Kong et al.}
\begin{document}

% \title{Discovery of a Quasar at $z=1.715$ with Blackbody Continuum of $T\sim10^4$~K: Transitioning from an LRD to a Normal Quasar?}

\title{A Wide and Deep Exploration of Radio-detected Active Galactic Nuclei with Subaru HSC (WERGS). XIII. \\
%High-Redshift Radio Quasar candidates beyond Ultra-Steep Spectrum Selection: Dropout selection from HSC--VLASS over $\sim$1200 deg$^2$}
High-z Radio Quasar Selection from HSC--VLASS over $\sim1200$ deg$^2$}

\correspondingauthor{Youwen Kong, Kohei Ichikawa}
\email{youwen.kong@ioa.s.u-tokyo.ac.jp, k.ichikawa@astr.tohoku.ac.jp}

\author[0009-0002-4201-7727]{Youwen Kong}
\affiliation{Institute of Astronomy, Graduate School of Science, The University of Tokyo, 2-21-1 Osawa, Mitaka, Tokyo, 181-0015 Japan}

\author[0000-0002-4377-903X]{Kohei Ichikawa}
\affiliation{Frontier Research Institute for Interdisciplinary Sciences, Tohoku University, Sendai, Miyagi 980-8578, Japan}
\affiliation{Astronomical Institute, Tohoku University, Aramaki, Aoba-ku, Sendai, Miyagi 980-8578, Japan}

\author{Hisakazu Uchiyama}
\affiliation{Department of Advanced Sciences, Faculty of Science and Engineering, Hosei University, Koganei, Tokyo 184-8584, Japan} 
\affiliation{National Astronomical Observatory of Japan, Mitaka, Tokyo 181-8588, Japan} 

\author[0009-0001-3910-2288]{Yuxing Zhong}
\affiliation{Department of Physics, School of Advanced Science and Engineering, Faculty of Science and Engineering, Waseda University, 3-4-1, Okubo, Shinjuku, Tokyo 169-8555, Japan}

\author[0000-0003-2682-473X]{Xiaoyang Chen}
\affiliation{Frontier Research Institute for Interdisciplinary Sciences, Tohoku University, Sendai, Miyagi 980-8578, Japan}

\author[0000-0002-4052-2394]{Kotaro Kohno}
\affiliation{Institute of Astronomy, Graduate School of Science, The University of Tokyo, 2-21-1 Osawa, Mitaka, Tokyo, 181-0015 Japan}
\affiliation{Research Center for the Early Universe, Graduate School of Science, The University of Tokyo, 7-3-1 Hongo, Bunkyo-ku, Tokyo 113-0033, Japan}
\affiliation{ILANCE, CNRS – University of Tokyo International Research Laboratory, Kashiwa, Chiba 277-8582, Japan}

\author{Tohru Nagao}
\affiliation{Research Center for Space and Cosmic Evolution, Ehime
University, 2-5 Bunkyo-cho, Matsuyama, Ehime 790-8577, Japan}

\author[0000-0003-4814-0101]{Kianhong Lee}
\affiliation{Department of Physics, Graduate School of Science, Nagoya University, Furo, Chikusa, Nagoya, Aichi 464-8602, Japan}

\author[0000-0003-2213-7983]{Bovornpratch Vijarnwannaluk}
\affiliation{Academia Sinica Institute of Astronomy and Astrophysics,
11F of Astronomy-Mathematics Building, AS/NTU, No.1, Section 4,
Roosevelt Road, Taipei 10617, Taiwan}

\author[0000-0001-5063-0340]{Yoshiki Matsuoka}
\affiliation{Research Center for Space and Cosmic Evolution, Ehime University, 2-5 Bunkyo-cho, Matsuyama, Ehime 790-8577, Japan}

\author[0000-0002-3531-7863]{Yoshiki Toba}
\affiliation{Department of Physical Sciences, Ritsumeikan University,
1-1-1 Noji-higashi, Kusatsu, Shiga 525-8577, Japan}
\affiliation{Academia Sinica Institute of Astronomy and Astrophysics,
11F of Astronomy-Mathematics Building, AS/NTU, No.1, Section 4,
Roosevelt Road, Taipei 10617, Taiwan}
\affiliation{Research Center for Space and Cosmic Evolution, Ehime
University, 2-5 Bunkyo-cho, Matsuyama, Ehime 790-8577, Japan}

\author[0000-0002-5956-8018]{Itsna Khoirul Fitriana}
\affiliation{National Astronomical Observatory of Japan, Mitaka, Tokyo 181-8588, Japan}
\affiliation{Astronomy Research Group, Institut Teknologi Bandung, Jl. Ganesha No. 10 Bandung 40132, Indonesia}

\author[0009-0007-6567-4240]{Sakiko Obuchi}
\affiliation{Department of Physics, School of Advanced Science and Engineering, Faculty of Science and Engineering, Waseda University, 3-4-1, Okubo, Shinjuku, Tokyo 169-8555, Japan}

\author[0009-0007-1714-5379]{Yuta Ishikawa}
\affiliation{Department of Physics, School of Advanced Science and Engineering, Faculty of Science and Engineering, Waseda University, 3-4-1, Okubo, Shinjuku, Tokyo 169-8555, Japan}

\author{Victor Kadri}
\affiliation{Ecole Polytechnique, Route de Saclay, 91128 Palaiseau cedex, France}
\affiliation{Institute of Astronomy, Graduate School of Science, The University of Tokyo, 2-21-1 Osawa, Mitaka, Tokyo, 181-0015 Japan}
\affiliation{ILANCE, CNRS – University of Tokyo International Research Laboratory, Kashiwa, Chiba 277-8582, Japan}




\begin{abstract}

We report the results of $g-$, $r-$, $i-$, and $z-$dropout selections
based on optical identifications of Very Large Array Sky Survey (VLASS) radio sources using the Hyper Suprime-Cam Subaru Strategic Program survey (HSC--SSP).
By positional crossmatching within $1\farcs5$ between the VLASS Epoch~2 catalog and the HSC--SSP Wide-layer catalog ($i \lesssim 26$), we obtain $\sim$400 high-redshift radio AGN candidates at $z \gtrsim 4$ over a $\approx1200~\mathrm{deg}^2$ survey footprint, extending optimistically to $z \sim 7$. 
%Optical magnitudes cluster at%
Their $i$-band AB magnitudes range from $i_\mathrm{AB} \simeq 24$--26, %indicating that these sources are largely inaccessible to shallower surveys such as SDSS.
placing most below the SDSS detection limit.
%By further crossmatching the HSC--VLASS dropout catalog with VLA Faint Images of the Radio Sky at Twenty-centimeters (FIRST) at 1.4~GHz, the LOFAR Two-metre Sky Survey (LoTSS) at 144~MHz, and the TIFR GMRT Sky Survey (TGSS) at 150~MHz, of which $\sim85\%$ have three radio detections,
%the majority of the high-$z$ candidates show flat to moderately steep radio spectra ($-1 \lesssim \alpha \lesssim 0$, with $f_\nu \propto \nu^\alpha$), and some also exhibit turnover radio spectra, demonstrating that conventional ultra-steep-spectrum (USS; $\alpha<-1.3$) selection would miss most of the population selected in this study.
By further crossmatching the HSC--VLASS dropout catalog with VLA Faint Images of the Radio Sky at Twenty-centimeters (FIRST) at 1.4~GHz, the LOFAR Two-metre Sky Survey (LoTSS) at 144~MHz, and the TIFR GMRT Sky Survey (TGSS) at 150~MHz, we obtain multi-frequency radio measurements for the majority of the sample. Approximately 85\% of the sources have three radio detections. The majority of the high-$z$ candidates show flat to moderately steep radio spectra ($-1 \lesssim \alpha \lesssim 0$, with $f_\nu \propto \nu^\alpha$), while some also exhibit turnover radio spectra, demonstrating a broad diversity of radio spectral properties among the dropout-selected candidates.
%We further perform SED fitting and obtain AGN luminosities, which %show a clustering at
%\rt{Further SED fitting derived AGN luminosities}
%are concentrated at
%typical bolometric luminosities of $\log(L_{\rm bol}/{\rm erg~s^{-1}})\sim46$--47. 
SED fitting yields typical AGN bolometric luminosities of
$\log(L_{\rm bol}/{\rm erg\,s^{-1}})\sim46.5$--47.5.
%We also examine the comoving number density distribution of our samples and find a sharp decline around the $i$-dropout regime ($z \sim 6$), suggesting the possible disappearance of luminous radio AGNs toward the epoch of reionization.
%\rt{We also examine the comoving number density distribution of our samples and find a declining trend from the $g$-dropout to the $z$-dropout regimes, with substantial decreases toward higher redshift, suggesting a declining population of luminous radio AGN candidates toward the epoch of reionization.}
We also find that the comoving number density declines from the
$g$-dropout to the $z$-dropout regimes, with substantial decreases
toward higher redshift, suggesting fewer luminous radio AGN candidates
toward reionization.
%We present an exploratory estimate of the observed comoving number density of the dropout-selected candidates, which declines toward higher redshift, although contamination and selection incompleteness prevent a robust determination of the intrinsic RLAGN number-density evolution.}

\end{abstract}




% Section 1
\section{Introduction}\label{sec:intro}

Understanding the role of active galactic nuclei (AGNs) in regulating star formation and driving the co-evolution of supermassive black holes (SMBHs) and their host galaxies is a central question in modern extragalactic astronomy. Tight empirical correlations between SMBH mass and host-galaxy properties indicate that SMBH growth and galaxy assembly are closely linked \citep{magorrian1998, ferrase2000, kormendy2013}. Among the various feedback channels, radio-mode feedback, mediated by relativistic jets launched from the vicinity of the SMBH, is thought to play a critical role in heating and redistributing the interstellar and circumgalactic medium, thereby regulating star formation and SMBH accretion over cosmic time \citep{best2005, fabian2012, heckman2014}.

At higher redshift ($z>2$), where both massive galaxies \citep[e.g.,][]{ilb13} and luminous AGNs  \citep[e.g.,][]{ued03,ued14,del14,air15,nii20,pou24}
become increasingly rare, characterizing the evolution of radio-loud AGNs (RLAGNs\footnote{
In general, the term ``radio-loud'' AGN refers to AGNs with either high radio loudness or powerful radio emission, typically defined by radio luminosities $L_{\mathrm{3\,GHz}} > 10^{24}$~W~Hz$^{-1}$. In this study, we use the term to denote AGNs with significant radio detections in the VLASS survey. %We note that the vast majority of our sources satisfy standard radio-loudness criteria, as discussed in Section~\ref{sec:RLfraction}.
}) is particularly important. This is especially true beyond the cosmic noon ($z \gtrsim 2$–3), where rapid changes in galaxy growth, black hole accretion, and large-scale structure formation are taking place \citep{behroozi2013,madau2014,he24}, extending toward the epoch of reionization ($z \sim 6$). Measuring the number density and physical properties of RLAGNs across this epoch provides a direct probe of jet-driven feedback in the early Universe \citep[e.g.,][]{vol11,obu26a,obu26} and its impact on massive galaxy formation.

%Radio-loud quasars and radio galaxies at high redshift represent key laboratories for such studies. Radio-loud quasars trace powerful jets interacting with the interstellar medium while remaining optically luminous, whereas radio galaxies, generally observed as Type 2 AGNs, are among the rarest populations at high redshift. High-redshift radio galaxies (HzRGs) are known to reside in exceptionally massive host galaxies \citep[$M_* \gtrsim 10^{11} M_\odot$;][]{seymour2007} and are frequently located at the centers of overdense environments, often associated with protoclusters traced by Ly$\alpha$ emitters (LAEs) and Lyman-break galaxies (LBGs) \citep{miley2008, overzier2016}. These systems are widely regarded as progenitors of today’s brightest cluster galaxies and massive galaxy clusters.

Radio-loud quasars and radio galaxies at high redshift represent key laboratories for studying the impact of powerful AGN jets on galaxy evolution \citep[e.g.,][]{nesvadba2017a,zhong2024}. Within the AGN unification framework, these populations are thought to be intrinsically similar systems hosting powerful radio jets, with their different observational properties arising primarily from orientation effects: radio-loud quasars are observed as unobscured (Type 1) AGNs with optically luminous nuclei, whereas radio galaxies are generally observed as obscured (Type 2) AGNs \citep[e.g.,][]{urr95}. Both radio populations are rare at high redshift and provide important probes of massive galaxy formation in the early Universe. High-redshift radio galaxies (HzRGs) are known to reside in exceptionally massive host galaxies \citep[$M_* \gtrsim 10^{11} M_\odot$;][Zhong et al. in prep.]{seymour2007} and are frequently located at the centers of overdense environments, often associated with protoclusters traced by Ly$\alpha$ emitters (LAEs) and Lyman-break galaxies (LBGs; \citealt{miley2008, overzier2016}). These systems are widely regarded as progenitors of today's brightest cluster galaxies and massive galaxy clusters.

%Historically, the search for HzRGs ($z \gtrsim 1$) has relied on the empirical correlation between redshift and radio spectral index, whereby high-redshift sources preferentially exhibit ultra-steep radio spectra (USS; $\alpha \lesssim -1.3$). This USS selection technique has proven effective in identifying distant radio galaxies and led to the discovery of the most distant confirmed HzRG to date, TN J0924–2201 at $z = 5.19$ \citep{debreuck2000, vanbreugel1999}. However, subsequent studies have demonstrated that USS selection is incomplete: a non-negligible fraction of $z \gtrsim 4$ radio galaxies and radio-loud quasars exhibit only moderately steep or even flat radio spectra and are therefore missed by USS-based searches \citep{jarvis2009, yamashita2020}.

Historically, searches for HzRGs at $z \gtrsim 1$ have relied on the empirical correlation between redshift and radio spectral index, where distant radio sources preferentially exhibit ultra-steep radio spectra (USS; $\alpha \lesssim -1.3$ with $f_\nu \propto \nu^\alpha$). Together with the well-established $K$--$z$ relation for radio galaxies, linking near-infrared $K$-band brightness to redshift \citep[e.g.,][]{willott2003}, the USS selection technique has proven effective in identifying HzRGs and led to the discovery of the most distant confirmed HzRG to date, TN J0924--2201 at $z = 5.19$ \citep{debreuck2000,vanbreugel1999}. These traditional approaches were primarily developed to identify radio galaxies, whereas high-redshift radio quasars are often preselected from optical or near-infrared surveys before being crossmatched with radio catalogs \citep[e.g.,][]{gloudemans2021,ighina2023}. Moreover, subsequent studies have shown that USS selection is incomplete, as a significant fraction of high-redshift radio sources exhibit only moderately steep spectra and would therefore be missed by USS-based searches \citep{jarvis2009,saxena2019,yamashita2020,zhong2023}. For example, radio-loud quasars frequently display flatter radio spectral indices ($-1 \lesssim \alpha \lesssim 0$), further limiting the effectiveness of USS-based selection for identifying the full population of high-redshift radio-loud AGNs.

One of the major challenges in identifying such high-redshift RLAGNs is the faintness of their optical and near-infrared emission \citep{debreuck2002}. Earlier searches using wide-area optical surveys, such as the Sloan Digital Sky Survey (SDSS; \citealt{abazajian2009}) in combination with the VLA-FIRST radio source catalog \citep{ivezi2002}, successfully identified optical counterparts over large sky areas and enabled extensive spectroscopic follow-up. %However, only $\sim30\%$ of the FIRST sources were matched with optical counterparts \citep{ivezi2002,helfand2015}. This low matching fraction is primarily due to the insufficient depth of SDSS, which limits its ability to robustly identify the faint optical counterparts of radio sources at higher redshifts.
However, only $\sim30\%$ of the FIRST sources were matched with optical counterparts \citep{ivezi2002,helfand2015}. This relatively low matching fraction is largely attributable to the limited depth of the SDSS imaging survey ($i_{\rm AB}\lesssim$22 mag), which restricts its ability to robustly identify the faint optical counterparts of radio sources at higher redshifts.

%A recent Subaru/Hyper Suprime-Cam \citep[HSC; ][]{miyazaki2018} strategic survey shed light on such a situation.
%The Wide and deep Exploration of Radio-detected active Galactic nuclei with Subaru/HSC (WERGS; \citealt{yamashita2018,tob19,ich21,ich23,uch22a,uch22b,uch22c,yam25})
%was initiated to construct a radio AGN and radio galaxy sample in the HSC SSP survey field covering the wide survey area reaching $\sim1200$~deg$^2$. 
A recent Subaru/Hyper Suprime-Cam \citep[HSC;][]{miyazaki2018} strategic survey has significantly improved the identification of faint optical counterparts to radio sources. The Wide and Deep Exploration of Radio-detected Active Galactic Nuclei with Subaru/HSC (WERGS; \citealt{yamashita2018,tob19,ich21,ich23,uch22a,uch22b,uch22c,yam25}) is a large-area catalog constructed by crossmatching deep HSC-SSP optical imaging with wide-area radio surveys, including VLA/FIRST and more recently VLASS \citep{uch26}, to establish a comprehensive sample of radio AGNs and radio galaxies over the HSC-SSP Wide layer, covering an area of $\sim1200$ deg$^2$.
%\cite{yamashita2018} demonstrated that over 60\% of populations now have reliable optical counterparts for VLA/FIRST radio sources thanks to deep HSC/optical imaging down to $i_\mathrm{AB}\approx26$, which is 3-4~mag deeper than the limits of earlier surveys such as SDSS. This WERGS catalog has uncovered rare and extreme systems, including radio-loud dust-obscured galaxies with prominent jet activity \citep{fuk25}, the discovery of a radio (and X-ray luminous) quasar at $z=3.4$ undergoing super-Eddington accretion \citep{obu26}.
\citet{yamashita2018} demonstrated that approximately 50\% of VLA/FIRST radio sources have reliable optical counterparts owing to the deep HSC imaging down to $i_\mathrm{AB}\approx26$, which is 3--4 mag deeper than the SDSS imaging survey and about 2--3 mag deeper than the Pan-STARRS1 (PS1) $3\pi$ survey ($i_{\rm AB}\lesssim23$; \citealt{flewelling2020}). More recently, optical and infrared counterpart identification rates have been pushed even higher with the LOFAR Two-Metre Sky Survey (LoTSS; \citealt{shimwell2022}), where 85\% of radio sources in Data Release (DR) 2 were assigned optical or infrared counterparts \citep{hardcastle2023}. The WERGS catalog has also uncovered rare and extreme systems, including radio-loud dust-obscured galaxies with prominent jet activity \citep{fuk25} and a radio- and X-ray-luminous quasar at $z=3.4$ undergoing super-Eddington accretion \citep{obu26a,obu26}.

One outstanding success of the WERGS survey was the discovery of the HzRG HSC J083913.17+011308.1 at $z = 4.72$, selected as an $r$-dropout and later spectroscopically confirmed \citep{yamashita2020}. 
%Notably, this source exhibits a moderately steep radio spectrum ($\alpha =-1.05$), flatter than the canonical USS criterion, and possesses a comparatively bright optical counterpart. 
%Similarly, several high-$z$ radio quasars have been discovered using optical dropout techniques \citep{banados2015,belladitta2020,gloudemans2023,ighina2025}, and these sources also exhibit non-USS-like radio spectra.
%These discoveries demonstrate that  combining optical–radio selection with dropout techniques can substantially reduce the selection bias inherent in USS-based searches,  enabling the identification of radio-loud AGNs spanning both quasar-like and radio galaxy-like populations and providing a more complete census of the high-redshift RLAGN population.
Notably, this source exhibits a moderately steep radio spectrum ($\alpha=-1.05$) and possesses a comparatively bright optical counterpart. Similarly, several high-redshift radio quasars have been discovered using optical dropout techniques \citep{banados2015,belladitta2020,gloudemans2023,ighina2025}, and these sources span a broad range of radio spectral properties. 
These discoveries demonstrate that combining optical--radio selection with dropout techniques enables the identification of RLAGNs spanning both quasar-like and radio-galaxy-like populations without imposing radio spectral-index pre-selection.

In this paper, we extend this methodology to a much wider area by crossmatching optical sources from the HSC-SSP with radio sources from the Very Large Array Sky Survey  \citep[VLASS;][]{lacy2020}. 
%This HSC–VLASS combination enables the construction of a large-area ($\approx1200$~deg$^2$), forming the final WERGS optical-identification catalog based on the completed HSC-SSP Wide survey \citep{uch26}. 
This HSC--VLASS combination enables the construction of a large-area ($\approx1200~\mathrm{deg}^2$) optical-identification catalog based on the completed HSC-SSP Wide survey, forming the final WERGS catalog \citep{uch26}.
By further incorporating ancillary radio information from FIRST and LoTSS, the resulting catalog provides homogeneous optical identifications, multi-frequency radio properties, and host-galaxy characterizations for more than $2\times10^4$ VLASS radio sources within the HSC footprint. %Building on this legacy dataset, we search for high-redshift AGNs at $z \gtrsim 4$ with substantially improved completeness compared to \rt{imposing radio spectral-index pre-selection.}
Building on this legacy dataset, we search for high-redshift AGNs at $z\gtrsim4$ without imposing radio spectral-index pre-selection.
%traditional USS-selected samples. 
%deep radio AGN sample and allows us to search for high-redshift AGNs at $z \gtrsim 4$ with significantly improved completeness.
%Our primary goals are to (i) recover high-redshift RLAGNs missed by previous USS-based selections, (ii) investigate the radio and host-galaxy properties of these sources, and %(iii) constrain the decline of the RLAGN number density toward the highest redshifts, where HzRGs become extremely rare.
%(iii) explore the observed distribution of dropout-selected RLAGN candidates toward the highest redshifts and compare it with previous studies.
Our primary goals are to 
(i) identify high-redshift RLAGNs spanning a broad range of radio spectral properties, 
(ii) investigate the radio and host-galaxy properties of these sources, and 
(iii) explore the observed number-density evolution of dropout-selected RLAGN candidates toward the highest redshifts and compare it with previous studies.

This paper is organized as follows. In Section \ref{sec:data}, we describe the optical and radio catalogs used in this study, along with additional radio surveys employed for crossmatching. Section \ref{sec:selection} details the selection procedures and criteria applied to construct the final source catalog. In Section \ref{sec:radio_optical}, we present the radio and optical properties of the candidate samples. In Section \ref{sec: AGN_host}, we present the crossmatched results with other surveys and the statistical properties of the host galaxies and AGNs in our sample. We discuss the implications for RLAGN populations and number density evolution in Section \ref{sec:RLAGN}. Our conclusions are summarized in Section \ref{sec:conclusion}. Throughout this paper, we assume a flat $\Lambda$CDM cosmology with $\Omega_m = 0.315$, 
$\Omega_\Lambda = 0.685$, and $H_0 = 67.4$ km s$^{-1}$ Mpc$^{-1}$ 
\citep{planck2020}.

% Section 2
\section{The Data}\label{sec:data}

The construction of the catalog follows a procedure similar to that of \citet{yamashita2018} and subsequent final WERGS catalog \citep{uch26}, in which radio AGNs are identified via positional crossmatching between the HSC--SSP optical source catalog and the VLASS Epoch~2 radio source catalog at 3\,GHz. The resulting HSC--VLASS crossmatched catalog is further cleaned to provide the parent sample for the selection of high-redshift candidates using the Lyman-break (dropout) technique based on their \textit{grizy} band colors.

\begin{figure}
    \centering
    \includegraphics[width=1.05\linewidth]{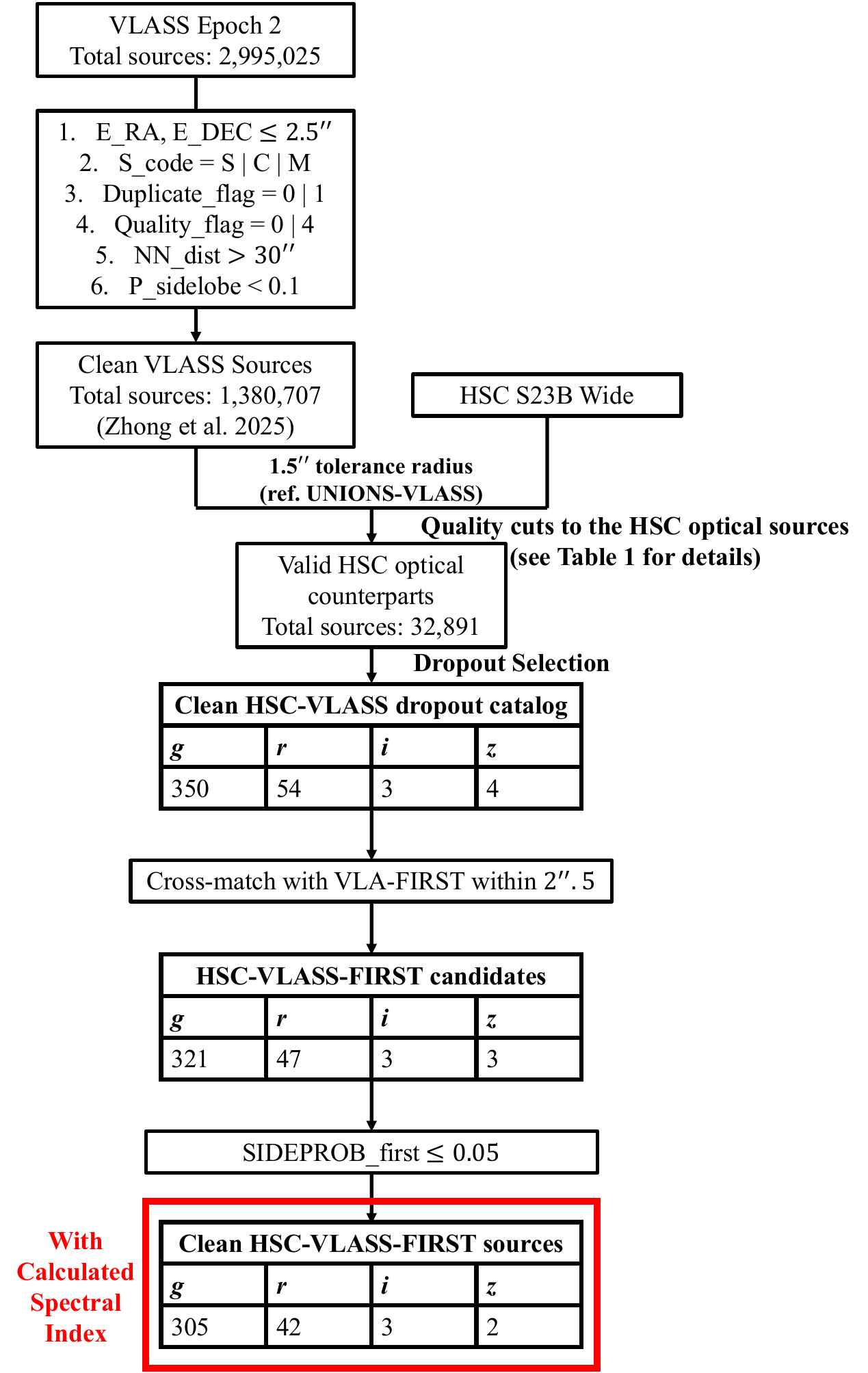}
    \caption{Flowchart illustrating the crossmatching process and the overall sample selection procedure. The detailed selection criteria used to validate the HSC optical counterparts are presented in Table~\ref{tab:hsc_flags} in Appendix~\ref{app:hscflag}.}
    \label{fig:flowchart}
\end{figure}

The final high-redshift candidate catalog serves as the reference sample and is subsequently crossmatched with additional radio surveys at different frequencies, including FIRST at 1.4\,GHz, TGSS at 150\,MHz, and LoTSS at 144\,MHz. Both radio source catalogs and image cutouts are utilized to characterize the multi-frequency radio properties of the selected sources. Figure~\ref{fig:flowchart} summarizes the full selection procedure, including the criteria applied for source cleaning and validation.

\begin{figure*}
    \centering
    \includegraphics[width=0.7\textwidth]{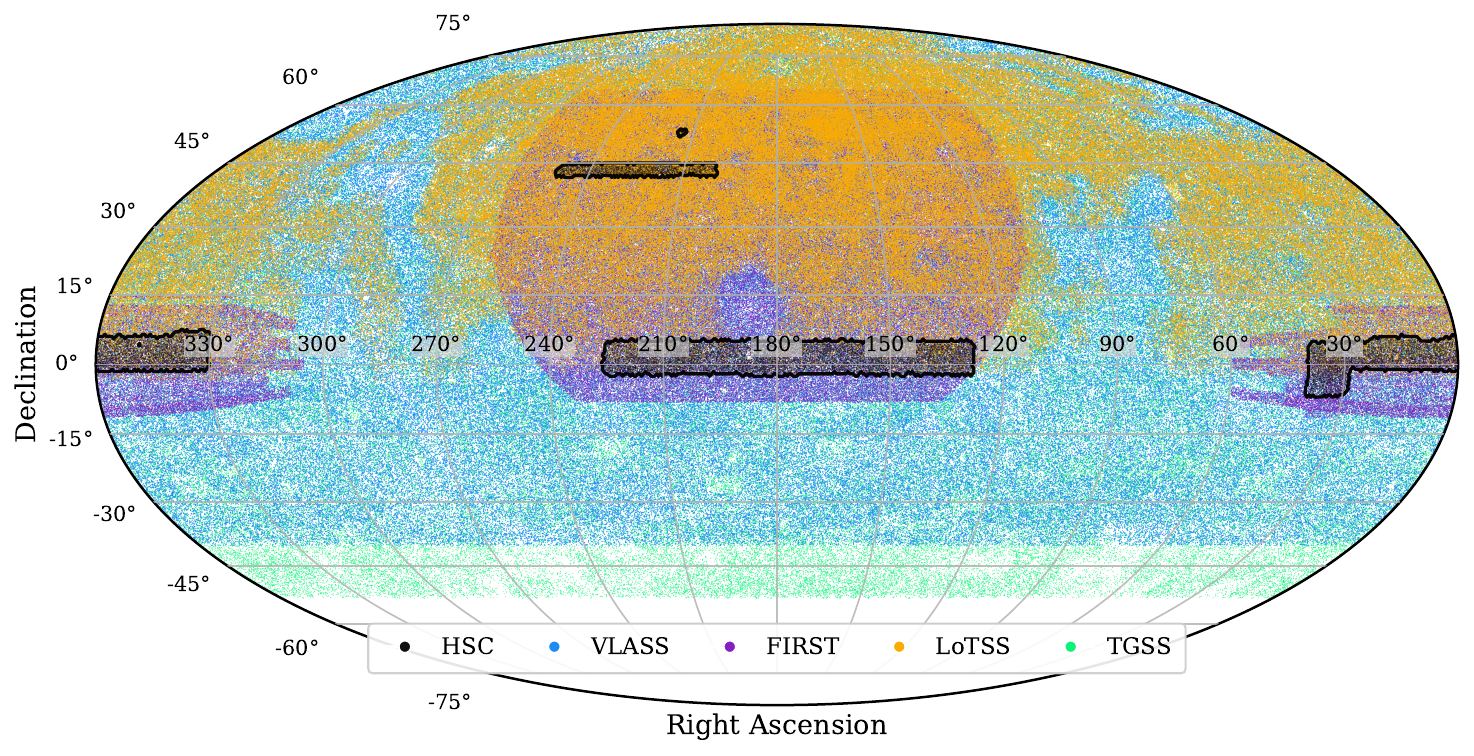}
    \caption{Sky coverage of the surveys used in this work: HSC Wide-area survey in optical (black), VLASS 3~GHz (blue), FIRST 1.4~GHz (purple), LoTSS 144~MHz (orange), and TGSS 150~MHz (green).}
    \label{fig:survey_coverage}
\end{figure*}
% frame for the HSC region

\subsection{HSC-SSP Final Data Release (S23B) Wide}

The basis of this study is the wide and deep optical imaging data from the Subaru Hyper Suprime-Cam Subaru Strategic Program (HSC-SSP). We use the internal final data release (S23B) of HSC-SSP, which includes observations obtained between 2014 March and 2022 January and represents the final data product covering the full HSC-SSP footprint \citep{aihara2022}.

The HSC-SSP survey consists of three layers with different combinations of area and depth (Wide, Deep, and UltraDeep). In this work, we exclusively utilize the Wide layer, which covers four distinct regions in the Northern Hemisphere: the Spring and Autumn Equatorial fields, HECTOMAP region at Dec $\sim$40 deg, and the AEGIS field, which is a single HSC pointing observed to the Wide layer depth. The total %effective survey area 
nominal survey footprint of the Wide layer is approximately $\sim1200~\mathrm{deg}^2$ (see Figure~\ref{fig:survey_coverage}).

The Wide layer provides deep optical imaging in five broad-band filters, $g$, $r$, $i$, $z$, and $y$, reaching typical $5\sigma$ point-source limiting magnitudes of $g\sim26.5$, $r\sim26.1$, $i\sim25.9$, $z\sim25.1$, and $y\sim24.4$ mag.
%$\sim26$~mag in the $g$, $r$, and $i$ bands and slightly shallower depths in $z$ and $y$. 
The median seeing of the Wide-layer data is $\sim0\farcs7$, enabling accurate source positions, reduced source confusion, and reliable photometric measurements.
%enabling reliable source detection, precise color measurements, and robust application of Lyman-break (dropout) selection techniques for high-redshift sources.

Photometric measurements are derived from aperture photometry using a $2''$ diameter aperture, with aperture corrections applied based on the point spread function (PSF).
%with corrections applied for point-spread-function effects. 
This photometric strategy provides consistent color measurements across the survey footprint and is well suited for identifying faint optical counterparts. %. 
A detailed description of the photometric measurements is given in Section ~\ref{sec:selection}.

\subsection{VLASS Epoch 2}
% what is the parent sample for VLASS epoch 2 (e.g. 3-sigma level or something), treatment for 'detection'.

The Very Large Array Sky Survey (VLASS; \citealt{lacy2020}) covers the sky at declinations $\mathrm{DEC} > -40^\circ$ with a total survey area of $33{,}885~\mathrm{deg}^2$ and covering the frequency range 2--4~GHz with a central observing frequency of 3~GHz, fully covering the HSC footprint. Observations are carried out primarily in the VLA B configuration, with the BnA configuration used to mitigate projection effects at low declinations, resulting in a typical angular resolution of $\sim2\farcs5$, significantly higher than the $5\farcs4$ resolution of FIRST. The single-epoch sensitivity reaches a 3$\sigma$ limit $\sim120\,\mu$Jy\,beam$^{-1}$, making VLASS one of the widest and most sensitive radio surveys at GHz frequencies.

VLASS data releases provide both Quick Look (QL) and Single-Epoch (SE) images. The QL images are produced using a rapid CLEAN process with a cell size of $1''$, while the SE images undergo deeper cleaning and self-calibration, achieving a finer cell size of $0\farcs6$. %As the SE images do not yet provide complete coverage of the HSC footprint, 
We adopt the SE products for positional crossmatching. Specifically, we use the Epoch~2 VLASS source catalog, which benefits from improved astrometric accuracy and mitigates the flux density underestimation issues present in the Epoch~1 release. The astrometric uncertainty is typically $\sim0\farcs5$ for sources near the catalog detection limit and improves to $\lesssim0\farcs3$ for bright compact sources.

For the best crossmatch performance while avoiding mismatches between optical and radio counterparts, we only consider compact radio sources %(i.e., isolated sources without nearby components within $3'$) 
for crossmatching with the HSC catalog. In practice, compact sources are selected using the VLASS clean-sample criteria, including \texttt{S\_code} $\neq$ E, \texttt{NN\_dist} $>30''$, and reliable astrometric and quality flags (see below), which preferentially retain single-component or simple Gaussian sources and exclude complex or extended radio morphologies. Therefore, before crossmatching with the HSC catalog, we apply an initial filtering of the VLASS catalog following the recommendations in the VLASS user guide \citep{gordon2021} and adopt the clean sample definition established in \citet{zhong2025}. Given the VLASS angular resolution of $\sim2\farcs5$, this corresponds to a physical scale of $\sim17$~kpc at $z\sim4$ (assuming the cosmology adopted in this work), and thus primarily selects compact radio AGN cores rather than large-scale jet structures. The applied criteria are:

\begin{enumerate}
    \item \texttt{E\_RA} and \texttt{E\_DEC} $\leq 2\farcs5$: astrometric uncertainties in both right ascension and declination are smaller than the VLASS angular resolution;

    \item \texttt{Duplicate\_flag} $< 2$: no duplicated source or duplicated counterpart within a search radius of $2''$;

    \item \texttt{Quality\_flag} $= (0|4)$: the source is either unflagged or has a peak flux density exceeding the integrated flux (\texttt{Peak\_flux} $>$ \texttt{Total\_flux});

    \item \texttt{S\_code} $\neq$ E: the source is well described by single- or multi-Gaussian components;

    \item \texttt{NN\_dist} $> 30''$: the angular separation between a VLASS source and its nearest neighbor in VLASS Epoch~2 exceeds $30''$, effectively removing complex radio structures (see Figure\,2 in \citealt{zhong2025});

    \item \texttt{P\_sidelobe} $< 0.1$: the probability that a source is a sidelobe artifact misclassified as a real detection is less than 0.1. This threshold corresponds to an estimated contamination rate of $\sim0.39\%$ and is adopted as the preferred cut. \citet{zhong2025} showed that applying the stricter criterion of \texttt{P\_sidelobe} $< 0.05$ does not significantly change the scientific results but would unnecessarily reduce the sample size. We therefore adopt the less restrictive threshold to preserve catalog completeness.
\end{enumerate}

We obtain a clean VLASS radio-source sample, reducing the total number of sources available for crossmatching to 1,380,707 (see the top three panels on the left of Figure~\ref{fig:flowchart}).

\subsection{FIRST}

The FIRST survey \citep{bec95,whi97} is a VLA 1.4\,GHz survey covering $8444~\mathrm{deg}^2$ in the North Galactic Cap and $644~\mathrm{deg}^2$ in the South Galactic Cap, with a declination coverage of $-10^\circ <\mathrm{DEC} < 65^\circ$ (Figure~\ref{fig:survey_coverage}). Its footprint fully overlaps the HSC survey area, enabling reliable source crossmatching and image retrieval for all HSC--VLASS sources.

We use the FIRST final data release source catalog \cite{helfand2015}, based on observations in the VLA B configuration, which provides a spatial resolution of $5\farcs4$ (FWHM) and a pixel size of $1\farcs8$. The typical rms noise is $\sim0.15$\,mJy\,beam$^{-1}$, with sources cataloged at approximately the $5\sigma$ level, corresponding to a flux density limit of $\sim1$\,mJy.
%The typical flux density limit of the survey is $\sim1$\,mJy.

\subsection{TGSS}

To investigate the multi-frequency radio spectral energy distributions of the selected radio sources, low-frequency radio data are required to complement the GHz-frequency surveys VLASS and FIRST. The TIFR GMRT Sky Survey (TGSS) is a 150\,MHz radio survey conducted with the Giant Metrewave Radio Telescope (GMRT; \citealt{intema2017}). The Alternative Data Release~1 (ADR1) covers nearly the entire sky north of $\mathrm{DEC} = -53^\circ$, with a median rms noise of $\sim3.5$\,mJy\,beam$^{-1}$. The corresponding source catalog contains more than 0.62 million sources detected above a $7\sigma$ threshold.

%Footprint of ADR1 covers the entire expected HSC-VLASS sky, however, due to its %relatively shallower depth, 
%shallow depth and low resolution of $25''$,}
%we primarily use the TGSS radio image cutouts for further check for complex radio sources, rather than relying on the catalog detection so that we retain as many candidates as possible for analysis.

The ADR1 footprint covers the entire HSC--VLASS survey area.
%However, because of its shallow depth and low angular resolution of $25''$, we primarily use the TGSS image cutouts to check for complex radio sources rather than relying on catalog detections, thereby retaining as many candidates as possible for further analysis.
 Given its relatively shallow depth and low angular resolution of $25''$, we inspect the TGSS image cutouts to identify potentially complex or confused radio sources in addition to using the catalog measurements.

\subsection{LoTSS}

Given the limited sensitivity of TGSS, we additionally make use of data from the LOFAR Two-metre Sky Survey DR~3 \citep{shimwell2026}
%Data Release~2 \citep{shimwell2022}
, which observes approximately %$27\%$ 
88\% of the northern sky at 144\,MHz with substantially improved sensitivity and angular resolution. %However, only the HECTOMAP and AEGIS HSC fields are currently covered by LoTSS DR2. 
The DR3 data products provide images at both $6''$ and $20''$ resolution, with the $6''$ mosaics reaching a median rms sensitivity of $\sim92 \mu$Jy beam$^{-1}$.
%We therefore use LoTSS image cutouts in conjunction with TGSS data to provide low-frequency radio constraints where available.
We use the public LoTSS DR3 Version~1.0 source catalog derived from the $6''$ mosaics and do not use the $20''$ products. To obtain the low-frequency radio measurements, we crossmatch the cleaned HSC--VLASS--FIRST sources with the TGSS ADR1 and LoTSS DR3 source catalogs using a matching radius of $2\farcs5$ and adopt the cataloged integrated flux densities. For both surveys, the image cutouts are used only to inspect complex or confused sources; no independent source fitting is performed.

\section{Sample Selection and Catalog Construction}
\label{sec:selection}

In this section, we describe the construction of the HSC--VLASS high-redshift radio AGN catalog and the procedure used to obtain multi-frequency radio information for the selected sources. The overall workflow is summarized in the flowchart shown in Figure~\ref{fig:flowchart}.

\subsection{Valid HSC Counterparts to VLASS 3~GHz Sources}

To optimally identify the true optical counterparts of radio sources while maintaining high completeness and minimizing contamination, we adopt a positional matching radius of $1\farcs5$ between HSC and VLASS sources. 
%This value follows the $1\farcs5$ crossmatch radius adopted for UNIONS--VLASS in \citet{zhong2025}, which yields a comparable contamination level and is therefore applicable to the HSC--VLASS dataset.
This choice follows the completeness--contamination analysis of \citet{zhong2025}, who found that a matching radius of $\sim1\farcs6$ provides a near-optimal balance, corresponding to a completeness of 92.7\% and a contamination rate of 7.5\% for matching VLASS sources to a deep optical survey. While \citet{uch26} subsequently adopted a more conservative $1\farcs0$ matching radius for the final WERGS catalog to further suppress chance HSC associations, we 
%retain the initial $1\farcs5$ radius 
used an initial crossmatching radius of $1\farcs5$
only for constructing the parent sample, thereby maximizing the completeness of the relatively small high-redshift candidate sample. The positional associations are subsequently refined through additional cleaning steps described below.

To construct a clean parent sample, we apply a set of quality cuts to the HSC optical sources following the HSC pipeline guidelines \citep{miyazaki2018}. These cuts are designed to remove spurious detections, saturated sources, and objects affected by image artifacts. The full set of criteria is summarized in Table~\ref{tab:hsc_flags} in Appendix~\ref{app:hscflag}.

We first select isolated or clearly deblended primary sources from the HSC detection catalog. We require that none of the central $3 \times 3$ pixels are saturated and that no bad pixels are present within the object footprint. In addition, we impose a minimum number of visits at the source position (\texttt{countinputs}) to ensure sufficient depth: at least three visits in the $g$ and $r$ bands and at least five visits in the $i$, $z$, and $y$ bands. Using this criterion, we obtain 32,891 valid HSC optical counterparts out of 36,654 initially matched VLASS radio sources.
%Using this criterion, we obtain the 32,891 valid HSC optical counterparts of the VLASS radio sources.

\subsection{LBG Selection}

To construct a parent sample suitable for Lyman-break galaxy (LBG) selection, we apply dropout-specific quality flags following \citet{ono2018} and \citet{harikane2022}. We require reliable flux measurements in the bands used for color selection, as summarized in Table~\ref{tab:hsc_flags}, %and enforce non-detection in the appropriate bluer bands for the $r$-, $i$-, and $z$-dropout samples. We also ensure that the photometry in the detection bands is not significantly affected by neighboring sources.
and ensure that the photometry in the detection bands is not significantly affected by neighboring sources.

\begin{figure*}
    \centering
    \includegraphics[width=\textwidth]{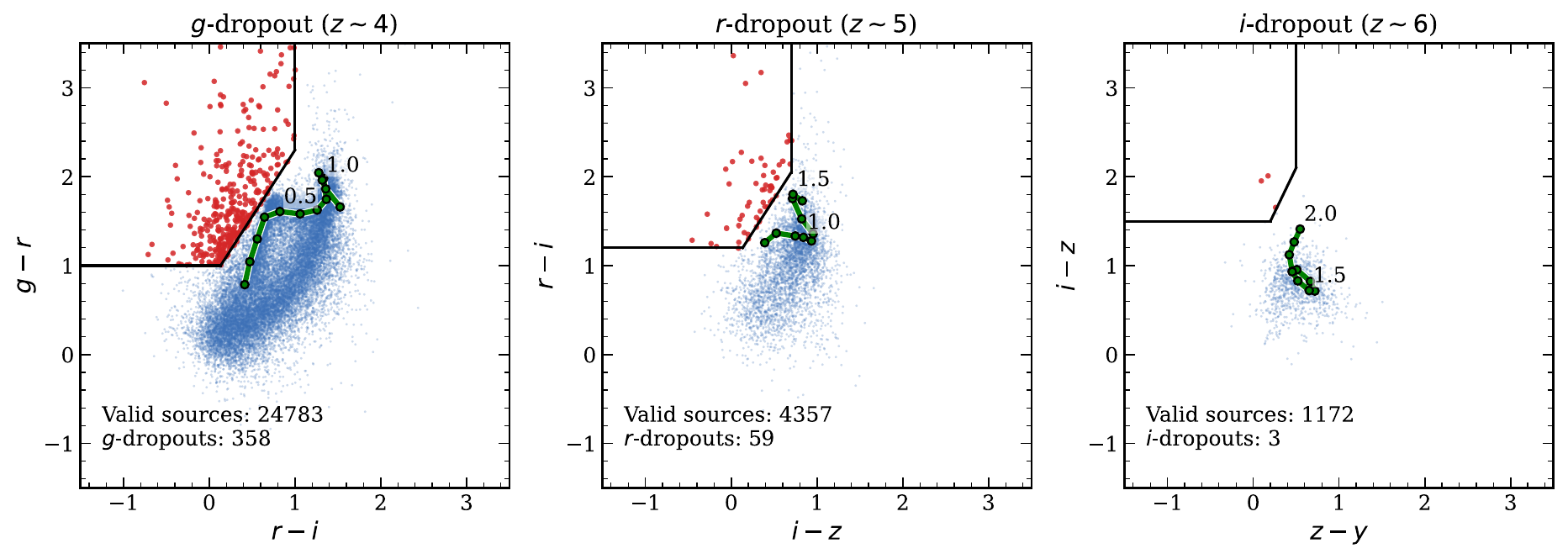}
    %\caption{Color--color diagrams used to select $g$-, $r$-, and $i$-dropout sources (left to right). The black lines indicate the adopted color-selection criteria (Equations~\eqref{equ:g_drop}, \eqref{equ:r_drop}, and \eqref{equ:i_drop}). Blue points show all sources in the clean HSC--VLASS parent sample, while red points mark the objects satisfying the corresponding dropout selection.}
    \caption{Color--color diagrams used to select $g$-, $r$-, and $i$-dropout sources (left to right). The black lines indicate the adopted color-selection criteria (Equations~\eqref{equ:g_drop}, \eqref{equ:r_drop}, and \eqref{equ:i_drop}). Blue points show all HSC--VLASS sources in the parent sample, while red points mark the selected dropout candidates. The green curves show evolutionary tracks of Balmer-break sources that may mimic the Lyman break and enter the selection regions, indicating possible low-$z$ contamination (see Appendix~\ref{app:low-z}). The numbers labeled along the green tracks indicate the corresponding redshifts of the Balmer-break sources.}
    \label{fig:color-color}
\end{figure*}

To construct the LBG-selected high-redshift candidate catalog, we use the forced photometry provided by the HSC pipeline, which measures fluxes in all bands using a consistent aperture defined in a reference band. By default, the $i$ band is adopted as the reference; for sources undetected in the $i$ or $z$ bands and bluer filters, the reference band is switched to $z$ or $y$, respectively. We then apply the color-selection criteria adopted from \citet{harikane2022} to identify LBG candidates at $z\sim4$, 5, 6, and 7 from the HSC multi-band photometric catalog, as summarized below.

\noindent
\textbf{$g$-dropouts ($z \sim 4$):}
\begin{equation}
    \begin{aligned}
       & g - r > 1.0,\\
       & r - i < 1.0,\\
       & g - r > 1.5(r - i) + 0.8;
    \end{aligned}
    \label{equ:g_drop}    
\end{equation}

\noindent
\textbf{$r$-dropouts ($z \sim 5$):}
\begin{equation}
    \begin{aligned}
        & r - i > 1.2, \\
        & i - z < 0.7, \\
        & r - i > 1.5(i - z) + 1.0;
    \end{aligned}
    \label{equ:r_drop}
\end{equation}

\noindent
\textbf{$i$-dropouts ($z \sim 6$):}
\begin{equation}
\begin{aligned}
    & i - z > 1.5,\\
    & z - y < 0.5,\\
    & i - z > 2.0(z - y) + 1.1;
\end{aligned}
    \label{equ:i_drop}
\end{equation}

\noindent
\textbf{$z$-dropouts ($z \sim 7$):}
\begin{equation}
    z - y > 1.6.
    \label{equ:z_drop}
\end{equation}

For reliable color measurements, \citet{harikane2022} noted that the latest HSC data release yields systematically brighter \texttt{cmodel} magnitudes compared to aperture-based measurements. We therefore compute source colors using magnitudes measured with a fixed $2''$ diameter aperture after aperture correction (\texttt{convolvedflux\_0\_20\_mag}), rather than the \texttt{cmodel\_mag} used in earlier work \citep[e.g.,][]{yamashita2018}. Accordingly, we also require that these fixed-aperture measurements are free of false flags. Using this procedure, we successfully identify dropout candidates from the HSC--VLASS crossmatched catalog (Figure~\ref{fig:color-color}).

\subsection{Clean HSC--VLASS Dropout Samples}

To construct the final clean HSC--VLASS dropout catalog, we first compute the angular separation between the HSC optical positions and the VLASS radio positions,which we record in the \texttt{HSC\_dist} column (see Table~\ref{tab: catalog_columns} in  Appendix~\ref{app:catalog_column}).

The initial HSC--VLASS dropout sample contains 358 $g$-dropouts, 59 $r$-dropouts, 3 $i$-dropouts, and 23 $z$-dropouts. We define sources with \texttt{HSC\_dist} $>1''$ as \texttt{on\_edge} candidates, corresponding to systems with relatively large optical--radio positional offsets. This criterion is adopted based on an examination of the full HSC--VLASS crossmatch results \citep{uch26}, which shows that the best balance between contamination and cleanliness is achieved at separations of $\sim1''$. 
%While a larger tolerance of $1\farcs5$ could retain a small number of sources with relatively large radio angular sizes, such cases represent only a minor fraction of the sample, and excluding them does not significantly affect the overall statistical results. All \texttt{on\_edge} sources are visually inspected using HSC image cutouts to verify the optical--radio associations. In cases where the matched optical source is clearly offset from the radio centroid and an alternative optical counterpart is found closer to the radio position, the source is classified as a mismatch and excluded from the sample. This step ensures reliable optical identifications for sources with larger positional offsets.
While the initial $1\farcs5$ matching radius is adopted to maximize the completeness of the parent sample, only a small fraction of candidates have separations exceeding $1''$. We therefore visually inspect all \texttt{on\_edge} sources using HSC image cutouts to verify the optical--radio associations. In cases where the matched optical source is clearly offset from the radio centroid and an alternative optical counterpart is found closer to the radio position, the source is classified as a mismatch and excluded from the sample. This step ensures reliable optical identifications for sources with larger positional offsets.

\begin{figure}
    \centering
    \includegraphics[width=1.05\linewidth]{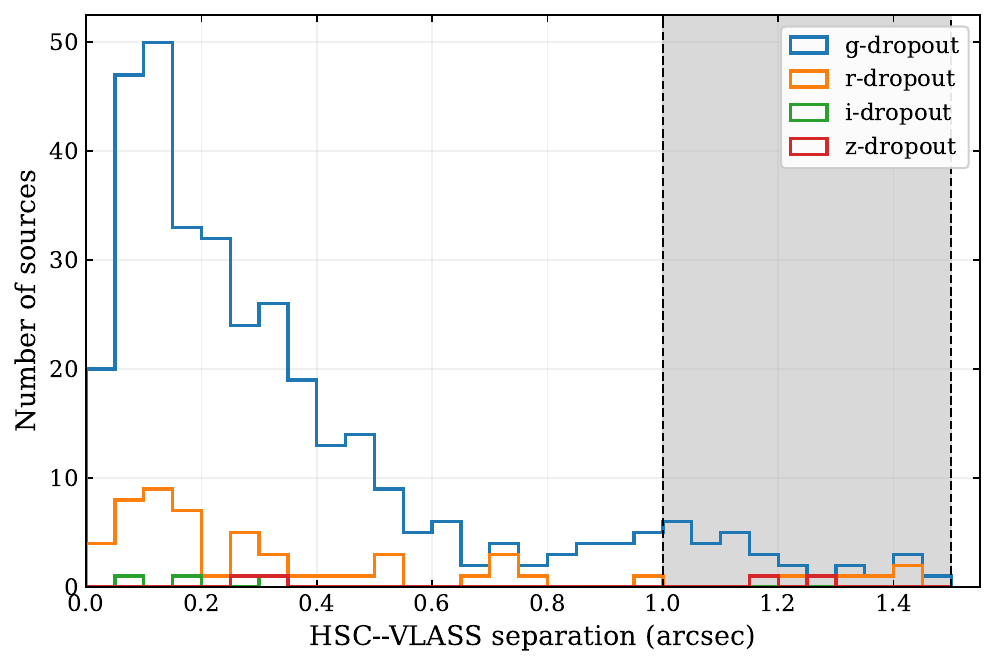}
    \caption{Distribution of the \texttt{HSC\_dist} between HSC optical positions and VLASS radio positions for the cleaned HSC--VLASS dropout samples. The shaded region marks sources with HSC--VLASS separations larger than $1''$, which are flagged as \texttt{on\_edge}.}
    \label{fig:hsc_dist}
\end{figure}

We next examine duplicate optical matches both within and across the different dropout categories. No duplicates are found across dropout groups, and only one duplicate match is identified within the $g$-dropout sample. For this case, we visually inspect the corresponding HSC image cutouts and compare their \texttt{HSC\_dist} values, retaining the association with the smaller offset and removing the counterpart with the larger separation, which is deemed the false match. Accounting for both large-offset mismatches and duplicate associations, a total of 8 sources are removed from the $g$-dropout sample, 5 from the $r$-dropout sample, and 12 from the $z$-dropout sample.

For the $z$-dropout sample, \citet{ono2018} noted that these sources are detected in only a single band, resulting in limited constraints in color--color space. We therefore perform an additional visual inspection of both coadded and warped HSC images for the remaining $z$-dropout candidates to remove spurious detections (e.g., apparent coadd detections caused by moving or transient objects in individual warps; see Appendix for examples). This step further excludes 7 $z$-dropout sources.

After applying these cleaning steps, we obtain a clean HSC--VLASS dropout catalog (%fourth panel of 
Figure~\ref{fig:flowchart}), consisting of 350 $g$-dropouts, 54 $r$-dropouts, 3 $i$-dropouts, and 4 $z$-dropouts. As shown in Figure~\ref{fig:hsc_dist}, although a small number of \texttt{on\_edge} sources remain, they account for only approximately 8.8\% of the final sample.
%their relative fraction is negligible.

As a final validation step, we crossmatch the cleaned dropout catalog with FIRST within $2\farcs5$ to obtain 1.4\,GHz flux measurements and to identify potential sidelobe-contaminated sources. %We apply the sidelobe validity criteria following \citet{yamashita2018} and remove candidates associated with likely sidelobe artifacts or complex radio structures.
We apply the sidelobe validity criteria following \citet{yamashita2018} and remove candidates associated with likely sidelobe artifacts or complex radio structures by requiring \texttt{P\_sidelobe} $< 0.05$, a conservative threshold adopted to minimize contamination from spurious FIRST detections.
%While the source counts do not change significantly after the positional crossmatch alone, the sidelobe-based cleaning further reduces the sample. We therefore adopt the cleaned HSC--VLASS--FIRST catalog as our final candidate catalog, yielding 305 $g$-dropouts, 42 $r$-dropouts, 3 $i$-dropouts, and 2 $z$-dropouts (bottom panel of Figure~\ref{fig:flowchart}).
The positional crossmatch reduces the source counts, with the $g$-dropout sample decreasing from 350 to 321 sources. The subsequent sidelobe-based cleaning further reduces the sample. We therefore adopt the cleaned HSC--VLASS--FIRST catalog as our final candidate catalog, yielding 305 $g$-dropouts, 42 $r$-dropouts, 3 $i$-dropouts, and 2 $z$-dropouts (bottom panel of Figure~\ref{fig:flowchart}).

\subsection{Samples with SDSS Spectroscopy}
\begin{figure}
    \centering
    \includegraphics[width=1.05\linewidth]{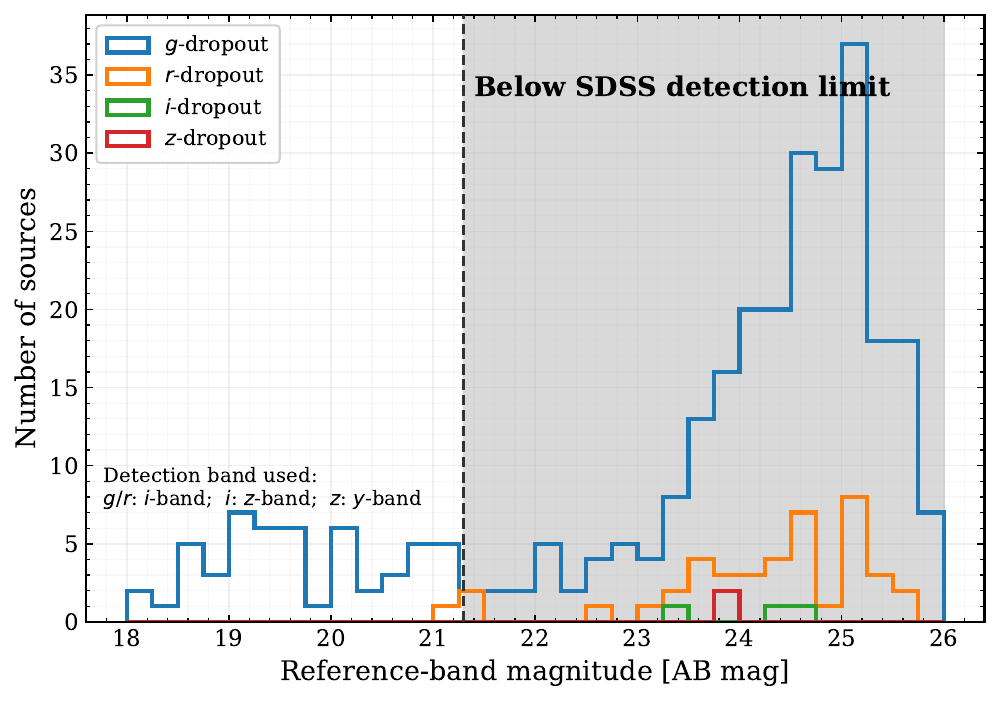}
    \caption{Distribution of reference-band magnitudes for the high-redshift radio-loud AGN candidates. The reference band denotes the detection band used for each source: $i$ for $g$- and $r$-dropouts, $z$ for $i$-dropouts, and $y$ for $z$-dropouts. The dashed vertical line indicates the representative SDSS $i$-band detection limit adopted for comparison, and the shaded region shows sources fainter than this limit. We note that the SDSS detection limits differ between bands; however, the $g$- and $r$-dropout samples, for which the $i$-band is used, dominate the sample, while only a few $i$- and $z$-dropout sources are included.}
    \label{fig:detection_mag}
\end{figure}
\begin{figure}
    \centering
    \includegraphics[width=1.05\linewidth]{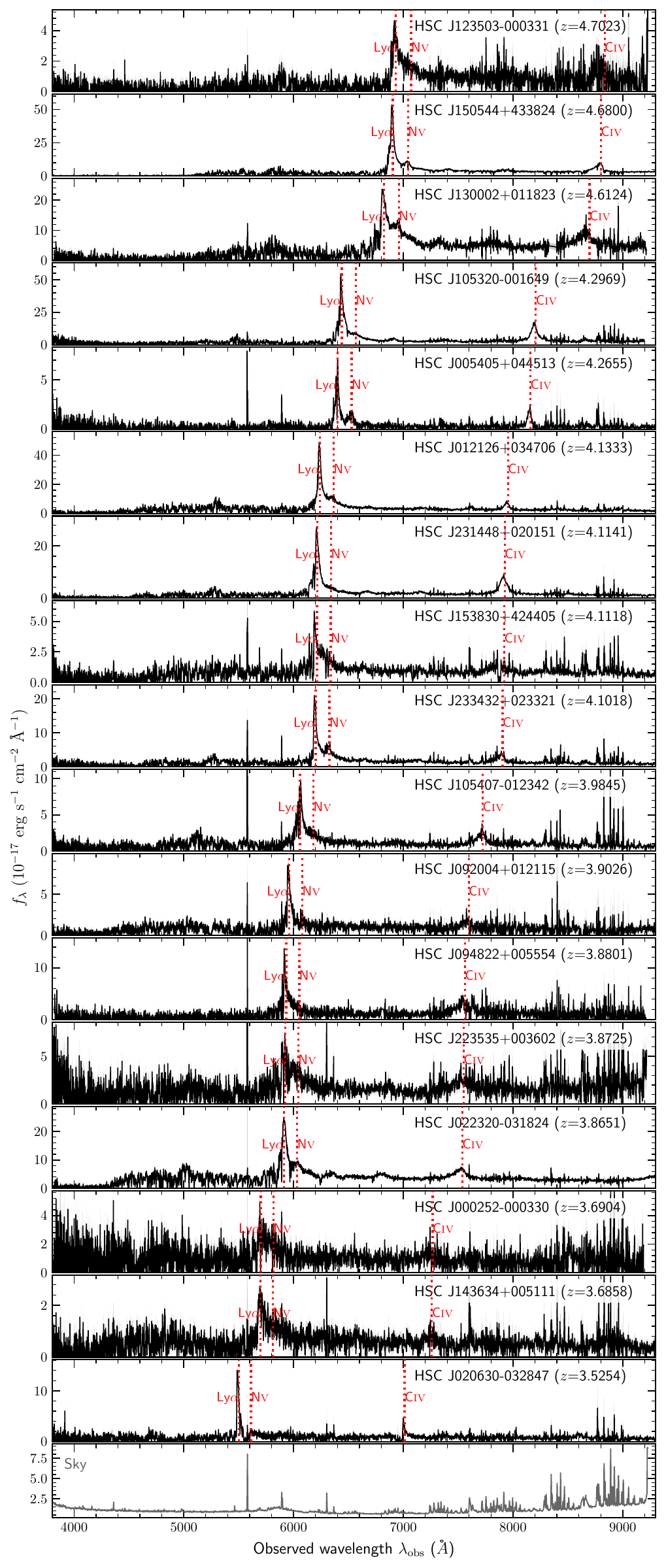}
    \caption{SDSS spectra of the 17 sources with spectroscopic redshifts consistent with their dropout colors.}
    \label{fig:sdss_spec}
\end{figure}

A subset of the relatively bright $g$-dropout candidates falls within the SDSS spectroscopic footprint, enabling an empirical consistency check of the dropout-based redshift selection.

We identify 37 sources from the $g$-dropouts with available SDSS spectroscopy. Among these, 29 spectra have sufficient quality (S/N $\gtrsim 3$) to reliably measure continuum and/or emission lines. Of the usable spectra, 17 objects have spectroscopic redshifts consistent with the expected redshift range of the $g$-dropout selection ($z\sim3.5-4.5$) (see Table~\ref{tab:sdss_source} in Appendix~\ref{app:sdssz}), and the remaining 12 are low-$z$ contaminants (see detailed discussion in Appendix~\ref{app:low-z}).
This corresponds to a spectroscopic confirmation fraction of $\sim60\%$, %broadly consistent with the success rates reported in previous HSC-based LBG studies \citep{ono2018,harikane2022}.
broadly consistent with the success rates reported for HSC-based $g$-dropout studies \citep{ono2018}. We note that this value is derived from the optically brightest subset of our sample with available SDSS spectroscopy and may therefore depend on source magnitude.
For the low-z interlopers, they are Balmer break sources at $z\sim0.5$, and we discuss such low-$z$ contaminants in Appendix~\ref{app:low-z}.

We emphasize that the SDSS spectroscopic sample is shallow and strongly biased toward optically bright sources ($i\lesssim21.5$ mag), while the majority of our candidates lie below the SDSS detection limit (Figure~\ref{fig:detection_mag}). Consequently, only a small fraction of the $g$-dropout sample has valid spectroscopy available (37 out of 305). For sources with consistent spectroscopic redshifts (Figure~\ref{fig:sdss_spec}), we adopt $z_{\rm spec}$ in subsequent SED fitting in place of photometric redshift estimates.

\section{Radio and Optical Properties} \label{sec:radio_optical}

\subsection{Radio loudness}
\label{sec:radio_loudness}

The conventional radio-loudness parameter is commonly defined using rest-frame radio and optical quantities (e.g., rest-frame 5~GHz and 4400~\AA\ flux densities; \citealt{kellermann1989}). Since our observations sample different rest-frame frequencies over the redshift range of the present sample, we apply a first-order $k$-correction to account for this difference.
We calculate the $k$-corrected radio loudness as

\begin{equation}
\begin{split}
\log_{10}R_{\rm kcorr}
&=
\log_{10}
\left(
\frac{F_{\rm radio}}
     {F_{\rm optical}}
\right)
+
\left(
\alpha_{\rm opt}
-
\alpha_{\rm radio}
\right)
\log_{10}(1+z)
\\
&=
0.4(m-t)
+
\left(
\alpha_{\rm opt}
-
\alpha_{\rm radio}
\right)
\log_{10}(1+z),
\end{split}
\label{equ:RKcorr}
\end{equation}
where $F_{\rm radio}\equiv F_{\rm int}$ is the total
(integrated) VLASS 3\,GHz flux density,
$F_{\rm optical}$ is the optical reference-band flux density,
$m$ is the corresponding optical magnitude, $t$ is the AB
magnitude corresponding to $F_{\rm radio}$, $z$ is the source
redshift, and $\alpha_{\rm radio}$ and $\alpha_{\rm opt}$ are
the radio and optical spectral indices, respectively. The
radio AB magnitude is given by

\begin{equation}
t~[{\rm AB~mag}]
=
-2.5
\log
\left(
\frac{F_{\rm radio}}
     {3631~{\rm Jy}}
\right).
\end{equation}
We adopt the $i$-band magnitude as the optical measurement for the $g$- and $r$-dropouts. For the $i$- and $z$-dropouts, we instead use the $z$- and $y$-band magnitudes, respectively, consistent with the corresponding reference bands.

We apply the first-order $k$-correction following the prescription adopted by \citet{ich21}. We adopt $\alpha_{\rm opt}=-0.5$ as the fiducial optical spectral index, while $\alpha_{\rm radio}$ is the representative radio spectral index adopted in Section~4.2. The resulting $k$-corrected radio loudness is denoted $R_{\rm kcorr}$ throughout this paper. Details of the correction and its dependence on the assumed optical spectral index are presented in Appendix~D.

\subsection{Radio Spectral Shape $\alpha$}\label{sec:radio_color}
\begin{figure}
    \centering
    \includegraphics[width=\linewidth]{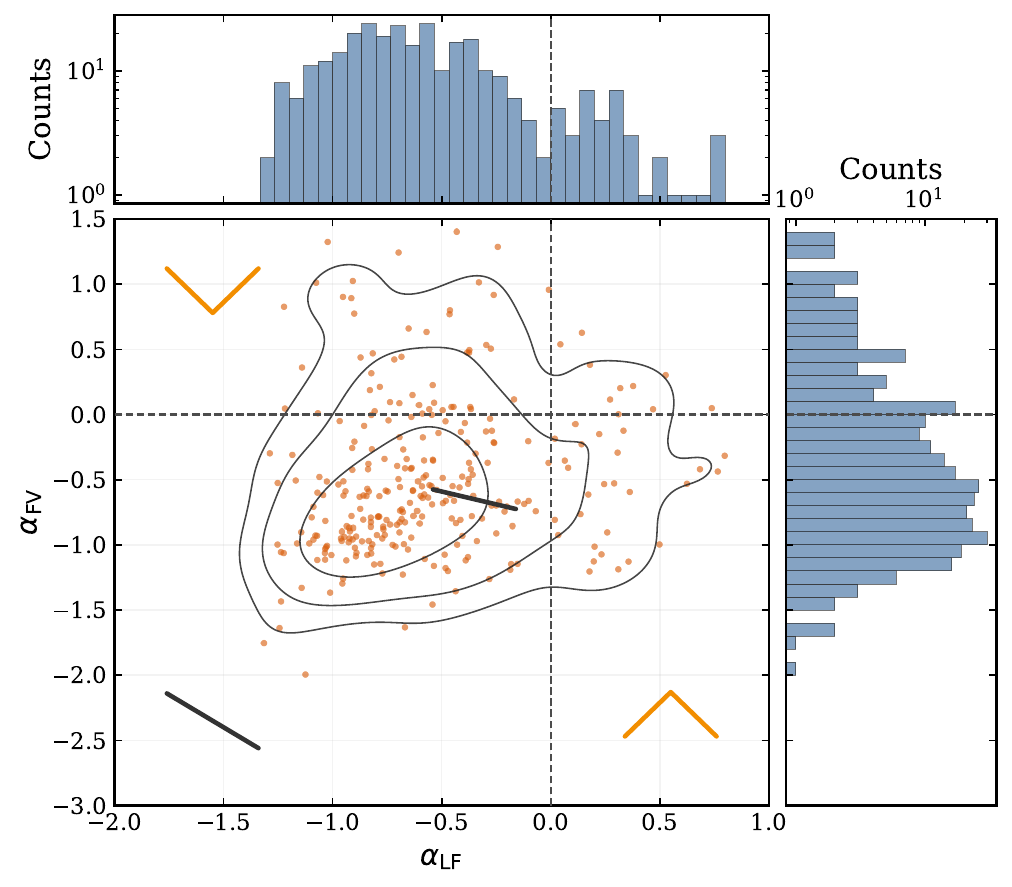}
    %\caption{Radio color--color diagram for sources with available low-frequency measurements, comparing the high-frequency spectral index $\alpha^{1.4\,{\rm GHz}}_{3\,{\rm GHz}}$ and the low-frequency spectral index $\alpha^{144/150\,{\rm MHz}}_{1.4\,{\rm GHz}}$. Marginal histograms show the corresponding distributions. The black dashed line indicates the cut between different radio spectral shapes, the red dashed line is the cut-off for USS at $\alpha^{1.4\,{\rm GHz}}_{3\,{\rm GHz}}=-1.3$ and $\alpha^{144/150\,{\rm MHz}}_{1.4\,{\rm GHz}}=-0.7$.}
    \caption{Radio color--color diagram for the 293 HSC--VLASS dropout-selected sources with available low-frequency measurements. The high-frequency spectral index, %$\alpha^{1.4\,{\rm GHz}}_{3\,{\rm GHz}}$
    $\alpha_{\rm FV}$, is plotted against the low-frequency index, %$\alpha^{144/150\,{\rm MHz}}_{1.4\,{\rm GHz}}$
    $\alpha_{\rm LF}$, with marginal histograms showing their distributions. %The red dashed lines mark the adopted USS-selection boundaries at %$\alpha^{1.4\,{\rm GHz}}_{3\,{\rm GHz}}
    %$\alpha_{\rm FV}=-1.3$ and %$\alpha^{144/150\,{\rm MHz}}_{1.4\,{\rm GHz}}
    %$\alpha_{\rm LF}=-0.7$, based on previous searches for high-redshift radio galaxies (e.g., \citealt{debreuck2000,singh2014}). The grey dashed lines at $\alpha_\mathrm{LF,FV}=0$ separate steep and inverted spectra. The shaded red and blue regions indicate the USS and flat-to-moderately steep regimes, respectively, while the schematic SEDs illustrate steep, moderately steep, peaked, and upturning radio spectra.
    The black contours indicate the kernel-density distribution of the sources. The grey dashed lines at $\alpha_{\rm LF,FV}=0$ separate positive and negative spectral slopes, while the schematic SEDs illustrate steep, moderately steep, peaked, and upturning radio spectra.}
    \label{fig:radio_color_color}
\end{figure}

The radio spectral shapes provide important diagnostics of the physical nature and evolutionary stage of %radio-loud AGNs (RLAGNs),
RLAGNs, as synchrotron emission dominates their radio spectral energy distributions (SEDs; e.g., \citealt{klamer2006}).
%Classical searches for high-redshift radio galaxies relied on USS selection, motivated by the empirical correlation between steep radio spectra and redshift (e.g., \citealt{debreuck2000,singh2014}), which may arise from a combination of enhanced inverse-Compton losses against the cosmic microwave background, spectral steepening toward higher rest-frame frequencies, and selection effects that preferentially identify steep-spectrum sources at high redshift \citep{miley2008,morabito2018}.
%Although USS selection is efficient for finding high-redshift radio galaxies, it does not provide a complete selection of radio AGNs. 
Some radio-loud quasars exhibit spectral curvature or turnover spectra (e.g., compact peaked-spectrum sources; \citealt{coppejans2015,shao2022,zhong2023}). This motivates the combination of VLASS (3\,GHz), FIRST (1.4\,GHz), and the low-frequency LoTSS (144\,MHz) and TGSS (150\,MHz) data, enabling both spectral-index measurements and searches for turnover radio SEDs.
%Consequently, a simple USS criterion does not recover the full population, particularly sources with moderately steep or flat radio spectra. 

%We characterize the HSC--VLASS dropout-selected RLAGNs using these multi-frequency data. Assuming $f_{\nu}\propto\nu^{\alpha}$,  we compute the radio spectral indices and first compute:
%(i) $\alpha_\mathrm{FV}$ between FIRST and VLASS (1.4--3~GHz) and $\alpha_\mathrm{LF}$ between low-frequency bands at 144/150~MHz and FIRST (144/150~MHz--1.4~GHz); and
%(ii) $\alpha_{\mathrm{LFV}}$ using VLASS, FIRST, and low-frequency flux densities when all three measurements are available.
%; and (ii) $\alpha_{\mathrm{FV}}$ from the FIRST--VLASS (1.4--3\,GHz) spectral slope for the remaining sources.
%Low-frequency measurements at 144/150\,MHz are available for 293 of the 352 sources. Among these, 76 are detected by both LoTSS and TGSS, for which we adopt the more sensitive LoTSS measurements.
%We then compute the representative spectral index adopted for the radio-luminosity calculations as

We characterize the HSC--VLASS dropout-selected RLAGNs using these multi-frequency data. Assuming $f_{\nu}\propto\nu^{\alpha}$, we compute $\alpha_{\mathrm{FV}}$ between FIRST and VLASS (1.4--3~GHz), $\alpha_{\mathrm{LF}}$ between the low-frequency bands at 144/150~MHz and FIRST (144/150~MHz--1.4~GHz), and $\alpha_{\mathrm{LFV}}$ using the VLASS, FIRST, and low-frequency flux densities when all three measurements are available. Low-frequency measurements at 144/150 MHz are available for 293 of the 352 sources. Among these, 76 are detected by both LoTSS and TGSS, for which we adopt the more sensitive LoTSS measurements. For the radio-luminosity calculations, we define the representative spectral index as

\begin{equation}
\label{equ: spec_index}
\alpha =
\begin{cases}
\alpha_{\mathrm{LFV}}, & \text{if $\alpha_{\mathrm{LFV}}$ is available},\\
\alpha_{\mathrm{FV}}, & \text{otherwise}.
\end{cases}
\end{equation}

We note that the spectral indices are derived from observed-frame frequencies and therefore probe different rest-frame frequency ranges depending on source redshift \citep[e.g.,][]{morabito2018}. Since the radio flux densities are compiled from surveys obtained at different epochs, intrinsic radio variability may introduce additional scatter in individual spectral indices \citep{barvainis2005,zhong2025}, although it is unlikely to affect the overall statistical trends.

Figure~\ref{fig:radio_color_color} shows a radio color--color diagram comparing the high-frequency spectral index $\alpha_\mathrm{FV}$ with the low-frequency spectral index $\alpha_\mathrm{LF}$. 
%The location of the sources in this plane provides a radio spectral curvature. 
%Most sources lie outside the classical USS region, defined here by $\alpha_{\mathrm{LF}}<-0.7$ and $\alpha_{\mathrm{FV}}<-1.3$ (red shaded region), and instead exhibit flat to moderately steep spectra.
Most sources exhibit flat to moderately steep spectra. In addition, approximately 30\% show spectral curvature. In particular, sources with $\alpha_{\mathrm{LF}}>0$ and $\alpha_{\mathrm{FV}}<0$ are consistent with spectra that peak between the low- and high-frequency bands.
%These results suggest that previous USS-based searches may have missed a substantial fraction of high-redshift RLAGNs, as our selection is sensitive to flat-spectrum and turnover-spectrum sources that would not be recovered by a simple USS criterion, including compact peaked-spectrum RLAGNs associated with young radio jets.
%We further discuss these implications in Section~\ref{sec:RLAGN}.
These results demonstrate the broad diversity of radio spectral shapes among the dropout-selected sources, including flat-spectrum and turnover-spectrum sources, as well as compact peaked-spectrum RLAGNs potentially associated with young radio jets.

% replace with spectral index vs. redshift (photo-z/spec-z/dropout colors if no) make two for 1.4/3 and 144/3 --> show the clustering using contours.--> emphasis that the previous study missed a lot of samples. --> expand the discussion for us picking up the intrinsically younger population --> different from previous studies. (uncertainties!)

%\section{AGN--host properties}
\section{Optical/IR AGN properties}
\label{sec: AGN_host}

\subsection{Multiwavelength photometry data}

The HSC footprint partially overlaps with several wide-area near-infrared (NIR) and mid-infrared (MIR) surveys, including VIKING \citep{edge2013}, UKIDSS LAS \citep{lawrence2013}, and AllWISE \citep{wright2019allwise}. These datasets provide additional NIR/MIR photometric constraints for a subset of our dropout-selected radio AGNs, particularly for relatively bright sources in the $g$- and $r$-dropout subsamples (e.g., $y < 22.5$\,mag).

We therefore crossmatch our catalog with VIKING and AllWISE using a search radius of $1\farcs5$ \citep[e.g.,][]{ich12,ich17} to obtain multiwavelength photometry where available. More than half of the $g$- and $r$-dropout sources have usable NIR and/or MIR counterparts, providing sufficient wavelength coverage for spectral energy distribution (SED) fitting and population-level analysis, as described in Section~\ref{sec:CIGALE}. 

\begin{figure*}
    \centering
    \includegraphics[width=0.6\linewidth]{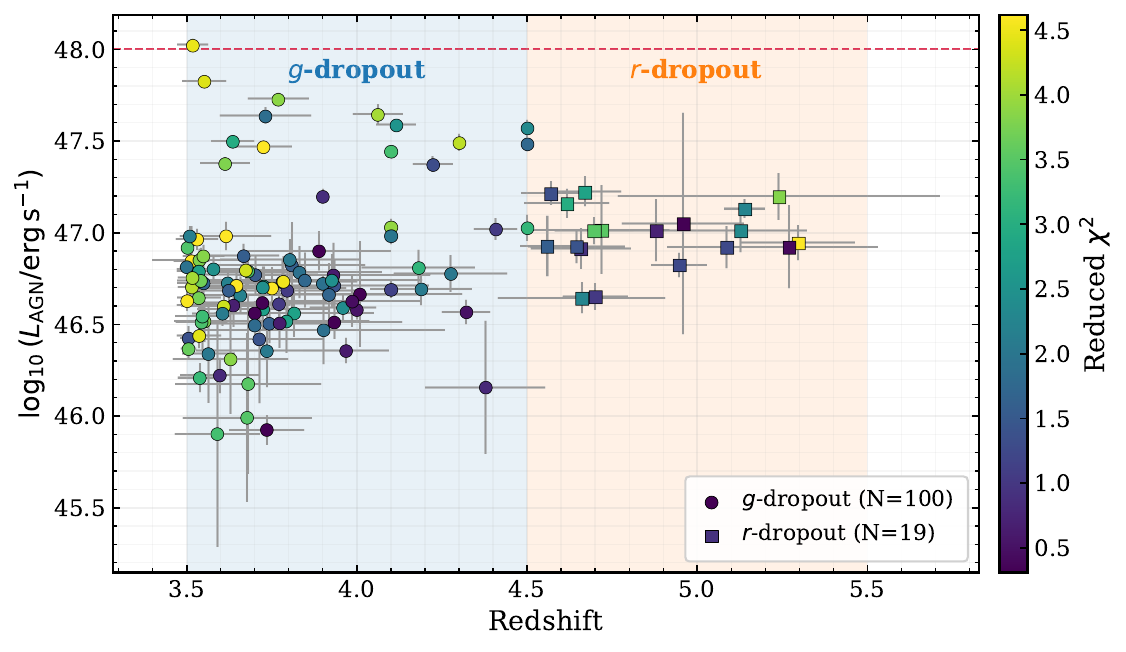}
    \caption{Luminosity--redshift ($L$--$z$) distribution of the $g$- and $r$-dropout radio-loud AGN sample derived from CIGALE SED fitting. Colors indicate the reduced $\chi^2$ of the best-fit model, and marker styles denote different dropout subsamples. The shaded regions indicate the redshift intervals adopted for the $g$- and $r$-dropout SED fitting. The red dashed line marks an approximate Eddington-limited bolometric luminosity of $\log_{10}(L_{\rm AGN}/\mathrm{erg\,s^{-1}})=48$, corresponding to a maximum BH mass limit of $M_{\rm BH}\sim10^{10}M_{\odot}$ \citep{ina16,ich17a}.}
    \label{fig:agn_lumi}
\end{figure*}

\subsection{CIGALE SED fitting}\label{sec:CIGALE}
% detail for CIGALE SED fitting to be put into the appendix, only the table of parameters...e.g what model is used...
To constrain AGN properties using the available multiwavelength photometry, we perform SED fitting with the Code Investigating GALaxy Emission (CIGALE; \citealt{boquien2019}). We fit all sources with at least one detection in VIKING or AllWISE, combining these measurements with the HSC optical photometry. Non-detections are included as upper limits following the standard CIGALE treatment. %For sources with SDSS spectroscopy, we fix the redshift to $z_{\rm spec}$ during the fitting. %In total, 148 $g$-dropouts and 26 $r$-dropouts satisfy these requirements and are included in the CIGALE SED fitting.
For sources with SDSS spectroscopy, we fix the redshift to $z_{\rm spec}$. In total, 148 $g$-dropouts and 26 $r$-dropouts satisfy these requirements.
None of the 3 $i$-dropouts or 2 $z$-dropouts has a sufficiently reliable VIKING or AllWISE counterpart for meaningful SED fitting, and these sources are therefore excluded from this analysis.

Motivated by the radio spectral index distribution in Section~\ref{sec:radio_optical}, which indicates that most objects exhibit flat-to-moderately steep spectra consistent with radio-loud quasars, we adopt a Type~1 AGN-dominated SED scenario. 
%This interpretation is further supported by the SDSS spectroscopic subsample, where all sources with redshifts consistent with the dropout selection show broad-line emission. 
This assumption is also supported by the SDSS spectroscopic subsample, in which the confirmed high-redshift sources show broad-line emission.
%In addition, the locations of the candidates in the color--color diagrams are consistent with the expected quasar evolutionary tracks at the corresponding dropout redshifts (Figure~\ref{fig:color-color}), providing independent support for the adopted redshift ranges.
For sources without spectroscopy, %we restrict the redshift range explored in the fitting to the dropout-inferred redshift intervals. 
%providing a physically motivated prior and reducing degeneracies in the photometric fits (See details in Appendix).
we assume that they lie within the dropout-inferred redshift intervals and restrict the redshift range explored in the fitting accordingly.

Since the primary goal of the SED fitting is to constrain the physical properties of the dropout-selected high-redshift candidates rather than to derive independent photometric redshifts, the dropout selection provides a physically motivated prior that reduces degeneracies in the photometric fits (see details in Appendix~\ref{app:cigale}). We note that a fraction of the fitted redshifts accumulate near the adopted dropout boundaries (e.g., $z\sim3.5$ and $z\sim4.5$), reflecting the imposed fitting priors and residual degeneracies in the available photometry. %Consequently, the fitted redshifts should be interpreted primarily as consistency checks with the dropout selection rather than as independent photometric-redshift determinations.
Consequently, the fitted redshifts are conditional on the adopted
dropout-based redshift ranges and should not be interpreted as
independent photometric-redshift determinations.

\subsection{AGN Luminosity}

Using the quasar-dominated SED model, we derive photometric redshifts and AGN bolometric luminosities for the $g$- and $r$-dropout subsamples. Figure~\ref{fig:agn_lumi} shows the resulting luminosity--redshift ($L$--$z$) distribution, where each source is color-coded by the reduced $\chi^2$ from the CIGALE fit.

To focus on robust fits, we apply a quality cut of $\chi^2_{\nu} \leq 5$, retaining 100 of the 148 fitted $g$-dropouts and 19 of the 26 fitted $r$-dropouts. 
The retained sources populate the luminous quasar regime, with typical values of $\log_{10}(L_{\rm AGN}/{\rm erg\,s^{-1}}) \sim 46.5$--47.5. This confirms that our dropout-selected radio sample preferentially recovers powerful AGNs at $z \gtrsim 4$, consistent with a population dominated by radio-loud quasars in the early Universe.

% background coloring for showing the range of g- and r-dropouts.
% figure for 3GHz luminosity vs. redshift (my fitted samples + Uchiyama's data)

\section{Redshift Evolution of Dropout-Selected Radio AGN Candidates}\label{sec:RLAGN}

\subsection{Radio Spectral Index vs. Redshift}\label{sec:spec_index}
\begin{figure*}
    \centering
    \includegraphics[width=0.6\linewidth]{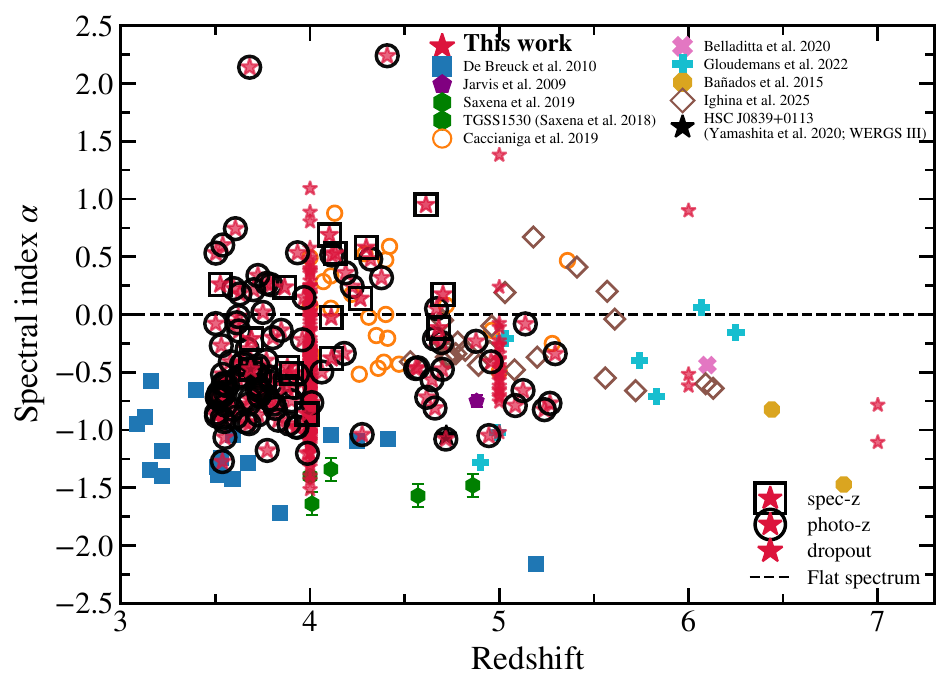}
    %\caption{Radio spectral index as a function of redshift for the sources in this work, compared with high-$z$ RLAGNs reported in the literature \citep{debreuck2010, jarvis2009, saxena2018, saxena2019, yamashita2020}. Red stars show our dropout-selected sample, with square outlines indicating spectroscopic redshifts, circular outlines indicating photometric redshifts, and filled stars indicating dropout-inferred redshifts. The colored symbols denote literature samples as indicated in the legend. The dashed horizontal line marks the flat-spectrum boundary, and the grey shaded region highlights the USS selection region ($\alpha \lesssim -1.3$).}
    %\caption{Distribution of radio spectral index as a function of redshift for the $z\gtrsim4$ radio-loud AGN candidates in this work, compared with high-redshift radio-loud AGNs reported in the literature \citep{debreuck2010,jarvis2009,saxena2018,saxena2019,yamashita2020,caccianiga2019,belladitta2020,gloudemans2023,banados2015,ighina2025}. Red stars show our dropout-selected sample, with square outlines indicating spectroscopic redshifts, circular outlines indicating CIGALE-derived photometric redshifts, and filled stars indicating dropout-inferred redshifts. The colored symbols denote literature samples as indicated in the legend. The dashed horizontal line marks the flat-spectrum boundary, while the grey shaded region highlights the conventional USS-selection regime ($\alpha \gtrsim -1.3$).}
    \caption{Distribution of radio spectral index as a function of redshift for the $z\gtrsim4$ radio-loud AGN candidates in this work, compared with high-redshift radio-loud AGNs reported in the literature \citep{debreuck2010,jarvis2009,saxena2018,saxena2019,yamashita2020,caccianiga2019,belladitta2020,gloudemans2023,banados2015,ighina2025}. Red stars show our dropout-selected sample, with square outlines indicating spectroscopic redshifts, circular outlines indicating CIGALE-derived photometric redshifts, and filled stars indicating dropout-inferred redshifts. The colored symbols denote literature samples as indicated in the legend. The dashed horizontal line at $\alpha=0$ separates positive and negative radio spectral slopes.}
    \label{fig:spectral_index}
\end{figure*}

%For a more comprehensive understanding of the spectral-index distribution, 

Figure~\ref{fig:spectral_index} compares the radio spectral index $\alpha$ (Section~\ref{sec:radio_color}) with the adopted redshift for the HSC--VLASS sample and previously identified high-$z$ radio AGNs.
 We use $z_{\rm spec}$ when available, the CIGALE-derived SED-based (photometric) redshift otherwise, and nominal dropout redshifts of $z=4, 5, 6$, and $7$ for the $g$-, $r$-, $i$-, and $z$-dropout samples, respectively, %when neither of other measurements is available. 
 when neither of the other measurements is available.
 For comparison, Figure~\ref{fig:spectral_index} includes
 %the conventional USS threshold of $\alpha\lesssim-1.3$ and high-$z$ radio AGN from the literature
high-$z$ radio AGNs from the literature
\citep{debreuck2010,jarvis2009,saxena2018,saxena2019}, including the dropout-selected radio galaxy HSC~J0839+0113 \citep{yamashita2020}.

Figure~\ref{fig:spectral_index} shows that no clear dependence of radio spectral index on redshift is apparent. Most HSC--VLASS sources have $-1\lesssim\alpha\lesssim0$ 
%and lie above the conventional USS threshold, 
consistent with the radio spectral indices reported for high-$z$ radio-loud quasars \citep{banados2015,caccianiga2019,belladitta2020,
gloudemans2023,ighina2025}. 
%Thus, the sample more closely resembles the high-$z$ radio-loud quasar population than the USS sources traditionally targeted in HzRG searches.
The broad range of spectral indices among both our candidates and previously identified high-redshift radio AGNs indicates substantial diversity in their radio spectral properties.

\subsection{3~GHz radio luminosity vs. Redshift}\label{sec:L3GHz_vs_z}

Using the radio spectral index $\alpha$ and redshift estimates defined in Section~\ref{sec:spec_index}, we compute the rest-frame 3\,GHz radio luminosity as

\begin{equation}
    L_{\nu} = \frac{4\pi D_L^2\, S_{\rm obs}}{(1+z)^{1+\alpha}},
\end{equation}
where $S_{\rm obs}$ is the observed flux density at frequency $\nu$, $D_L$ is the luminosity distance, and $\alpha$ is the radio spectral index. This expression accounts for the standard $k$-correction assuming a power-law radio spectrum.
For sources without spectroscopic or photometric redshifts, the luminosities should be regarded as approximate population-level estimates rather than precise measurements for individual objects.

\begin{figure*}
    \centering
    \includegraphics[width=0.7\linewidth]{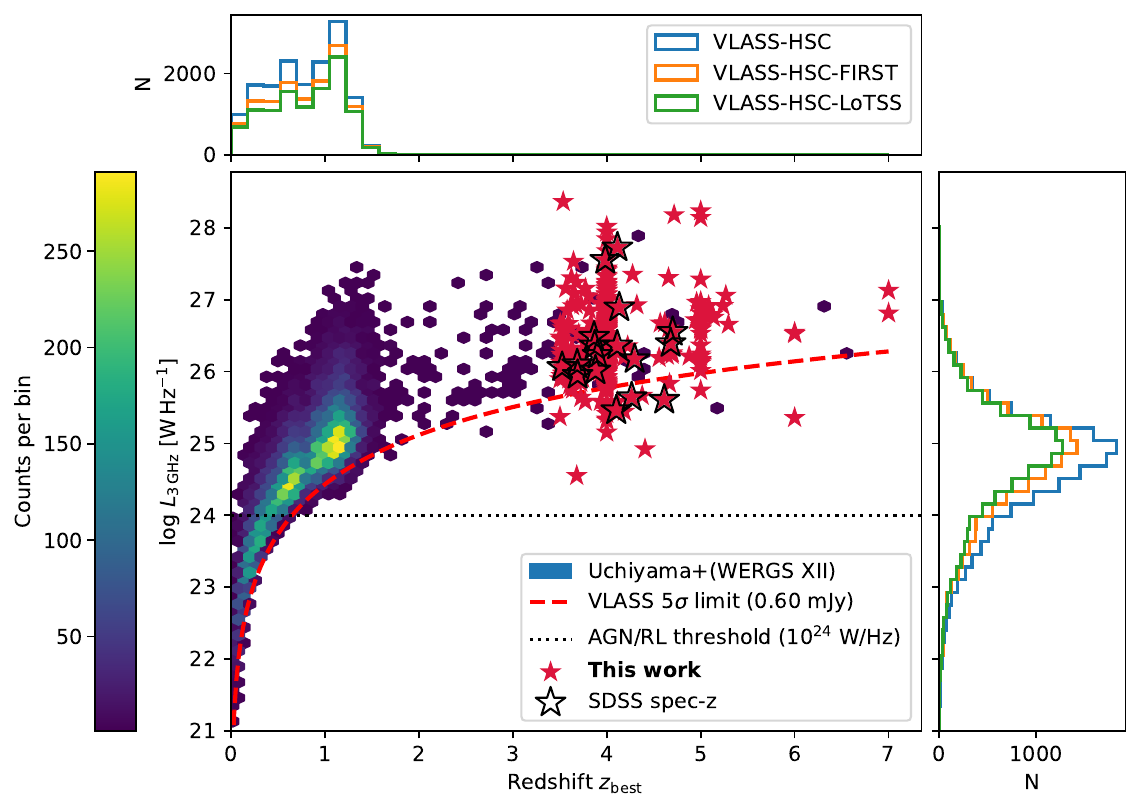}
    %\caption{Comparison of our dropout-selected sources (red stars) with the full HSC--VLASS radio AGN population (density map) and the VLASS $5\sigma$ sensitivity limit (Uchiyama et al.\ in prep.), together with the radio-loud AGN threshold.}
    \caption{Comparison of our dropout-selected radio-loud AGN candidates (red stars) with the full HSC--VLASS radio AGN population \citep[density map;][]{uch26} in the redshift--radio luminosity plane. Black-outlined stars denote sources with SDSS spectroscopic redshifts. The red dashed curve shows the VLASS $5\sigma$ sensitivity limit. %while the black dotted line indicates the commonly adopted radio-power reference of $L_{3\,\mathrm{GHz}} = 10^{24}\,\mathrm{W\,Hz^{-1}}$.
    The black dotted line marks $L_{3\,\mathrm{GHz}} = 10^{24}\,\mathrm{W\,Hz^{-1}}$, following the radio-power reference adopted by \citet{uch26}; it is shown here to facilitate direct comparison with the parent HSC--VLASS population and is not used as a radio-loudness classification criterion in this work.}

    \label{fig:uchiyama_overlay}
\end{figure*}

Figure~\ref{fig:uchiyama_overlay} shows the $L_\mathrm{3\,GHz}$ distribution as a function of redshift of our HSC--VLASS drop out sources, as well as the full HSC--VLASS radio AGN sample constructed by \citealt{uch26}. 
Our sources occupy the high-redshift extension of the broader radio AGN distribution. The SDSS spectroscopically confirmed sources span a similar radio-luminosity range to the overall candidate sample.

A small fraction of sources fall below the nominal VLASS $5\sigma$ sensitivity limit. This may reflect uncertainties in the adopted redshifts and radio spectral indices used for the $k$-correction, differences between the nominal sensitivity curve and the effective survey depth, or contamination by lower-redshift sources. Nevertheless, the majority of the candidate sample occupies the region of the redshift--radio luminosity plane expected for luminous high-redshift radio AGNs. Spectroscopic follow-up is required to confirm individual candidates, particularly at higher redshift where contamination may increase, and to refine the derived luminosity distributions.

\subsection{Dropout-based comoving number density of RLAGNs}\label{sec:number_density}
\begin{figure}
    \centering
    \includegraphics[width=\linewidth]{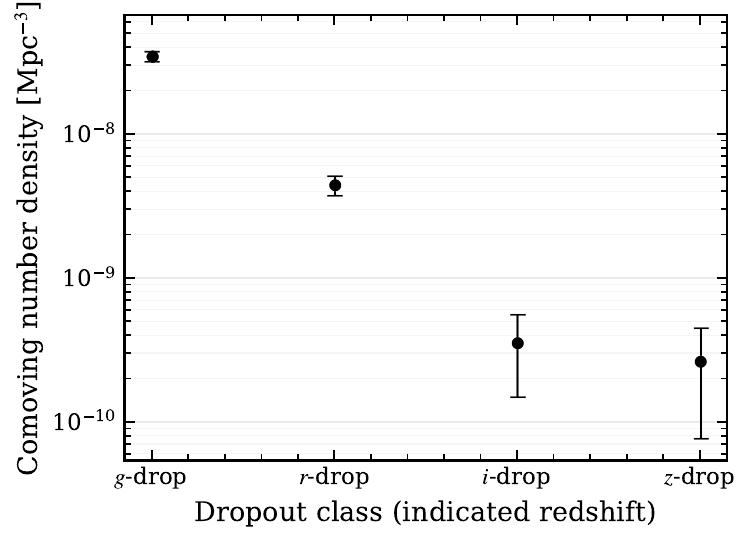}
    \caption{Dropout-based comoving number density of RLAGNs in the HSC--VLASS sample. A uniform fiducial contamination fraction of 40\% is applied to all dropout samples. Error bars indicate Poisson uncertainties after applying the same correction.}
    \label{fig:number_density}
\end{figure}

We estimate the number density of RLAGNs at $z\sim4$--7 using a simplified effective-volume approach. The effective survey area of the HSC S23B Wide layer is estimated after masking bright stars and low-quality regions \citep[see][]{harikane2022}, resulting in an unmasked area of $\approx551.675~\mathrm{deg}^2$ \citep{uch26}. As a first-order approximation, we assume that dropout-selected sources lie within the corresponding color-selected redshift intervals for each dropout class. Under this assumption, the effective comoving survey volume \citep{steidel1999} is computed as

\begin{equation}
V_{\rm eff} = A_{\rm eff} \int_{z_1}^{z_2} \frac{dV}{dz\,d\Omega}\,dz,
\label{equ:effective_area}
\end{equation}
where $A_{\rm eff}$ is the effective solid angle and $z_1$--$z_2$ represent the redshift range associated with each dropout selection.

In principle, the comoving number density corrected for contamination can be written as

\begin{equation}
\psi_{\rm bin} = \frac{1}{V_{\rm eff}} \sum_m \left[1-f_{\rm cont}(m)\right] N_{\rm raw}(m).
\label{equ:bin_sum}
\end{equation}

\begin{equation}
\psi(m) = \left[1-f_{\rm cont}(m)\right]\frac{N_{\rm raw}(m)}{V_{\rm eff}},
\label{equ:correction}
\end{equation}
%where $N_{\rm raw}(m)$ is the number of RLAGN candidates in each magnitude bin and $f_{\rm cont}(m)$ is the contamination fraction adopted from \citet{ono2018}. 
where $N_{\rm raw}(m)$ is the number of RLAGN candidates in each magnitude bin and $f_{\rm cont}(m)$ represents the magnitude-dependent contamination fraction.
Such corrections are commonly adopted when deriving dropout luminosity functions  \citep{ono2018} and are applied in Section~\ref{sec:lumi_func}.

For the present analysis, we estimate the overall comoving number density of dropout-selected RLAGN candidates within each dropout redshift interval rather than a magnitude-dependent luminosity function. We therefore adopt a uniform fiducial contamination fraction of $f_{\rm cont}=0.40$ for all four dropout samples and use the estimator

%For simplicity, and because detailed completeness simulations are not available for our radio-selected sample, we adopt the uncorrected estimator

\begin{equation}
\psi_{\rm dropout}=
\left(1-f_{\rm cont}\right)
\frac{N_{\rm raw}}{V_{\rm eff}},
\qquad
f_{\rm cont}=0.40,
\label{equ:density}
\end{equation}
with Poisson uncertainties
\begin{equation}
\sigma
=
\left(1-f_{\rm cont}\right)
\frac{\sqrt{N_{\rm raw}}}{V_{\rm eff}}.
\label{equ:density_err}
\end{equation}

%We note that this approach does not account for contamination within the dropout-selected sample. Spectroscopic follow-up of the current catalog suggests that the contamination fraction is $\sim40\%$. 
%and is likely to increase toward higher-redshift dropout selections. 
%Applying such a correction would systematically reduce the inferred number densities, but a robust contamination model for the radio-selected sample is beyond the scope of this work.}
We do not apply the magnitude-dependent contamination corrections of \citet{ono2018}, because they were derived for optically selected HSC dropout samples rather than radio-selected RLAGN candidates. The bright SDSS spectroscopic subsample suggests a low-redshift contamination fraction of approximately 40\%. We therefore adopt this value as a uniform fiducial correction for all four dropout samples. This correction remains approximate because a representative contamination model for the full radio-selected sample is not yet available.

%Spectroscopic follow-up of the current catalog indicates that the low-redshift contamination fraction is approximately $\sim40\%$. Applying such a correction would systematically reduce the absolute normalization of the inferred number densities, but a robust contamination model for the radio-selected sample is not yet available and is beyond the scope of this work.}
 
Figure~\ref{fig:number_density} shows the results computed using Equations~(\ref{equ:effective_area}), (\ref{equ:density}), and (\ref{equ:density_err}).
%We note that this estimate does not take into account selection completeness or redshift-dependent selection effects inferred from apparent magnitude, as implemented in \citet{ono2018}, and should therefore be interpreted only as an approximate indication of number-density evolution across the dropout bins.
%This estimate does not account for selection completeness, contamination, or magnitude- and redshift-dependent selection effects. The derived number densities should therefore be regarded as uncorrected estimates and interpreted only as a preliminary description of the observed candidate distribution across the dropout bins.}
This estimate does not account for selection completeness or magnitude- and redshift-dependent selection effects. The derived number densities include only the uniform fiducial 40\% contamination correction and should therefore be interpreted as a preliminary description of the candidate distribution across the dropout bins.

%The inferred number density declines from $z\sim4$ to $z\sim7$, with a particularly sharp drop between the $r$-dropout ($z\sim5$) and $i$-dropout ($z\sim6$) bins. Even with this inclusive estimate, we identify a small number of candidates beyond $z\sim6$. 
%However, the present measurements remain strongly affected by contamination and selection incompleteness, both of which are expected to become increasingly important toward higher redshift.}
%This decline may reflect an intrinsic decrease in the space density of luminous radio-loud AGNs at very high redshift, but it may also be amplified by selection effects and radio detectability limitations, including enhanced inverse-Compton cooling off the cosmic microwave background \citep{mocz2011, ghisellini2014}, which suppresses synchrotron emission from extended radio structures (``CMB quenching'').
%The inferred candidate number density declines from the $g$-dropout to the $z$-dropout bin, with a particularly sharp drop between the $r$-dropout ($z\sim5$) and $i$-dropout ($z\sim6$) samples.
The inferred candidate number density declines from the
$g$-dropout to the $z$-dropout bin, with substantial decreases between
both the $g$- and $r$-dropout and the $r$- and $i$-dropout samples.
A small number of candidates are nevertheless identified beyond $z\sim6$. Both the absolute normalization and the relative trend remain uncertain because contamination and selection incompleteness may vary among the dropout classes. The result should therefore be interpreted as an observed decline in the dropout-selected candidate density, rather than a direct measurement of the intrinsic RLAGN number-density evolution.

%The inferred candidate number density declines from the $g$-dropout to the $z$-dropout bin, with a particularly sharp drop between the $r$-dropout ($z\sim5$) and $i$-dropout ($z\sim6$) samples. A small number of candidates are nevertheless identified beyond $z\sim6$. Both the absolute normalization and the relative trend remain uncertain because the actual contamination fraction and selection incompleteness may vary among the dropout classes. The result should therefore be interpreted as an observed decline in the dropout-selected candidate density, rather than a direct measurement of the intrinsic RLAGN number-density evolution.}

An intrinsic decline in the space density of luminous AGNs beyond the peak of quasar activity at $z\sim2$--3 is expected from previous studies \citep{hopkins2007}.
%However, the current data do not allow us to disentangle such intrinsic evolution from observational selection effects. In particular, the apparent decline between the $r$- and $i$-dropout bins is likely influenced by increasing incompleteness and contamination in the dropout selection. 
Although the observed decrease is qualitatively consistent with this trend, the current data cannot distinguish intrinsic evolution from residual selection effects. Enhanced inverse-Compton losses against the cosmic microwave background \citep{mocz2011,ghisellini2014} may further reduce the detectability of extended radio emission at high redshift, although their contribution to the observed decline cannot be quantified with the present sample.

\begin{figure*}
    \centering
    \includegraphics[width=0.6\linewidth]{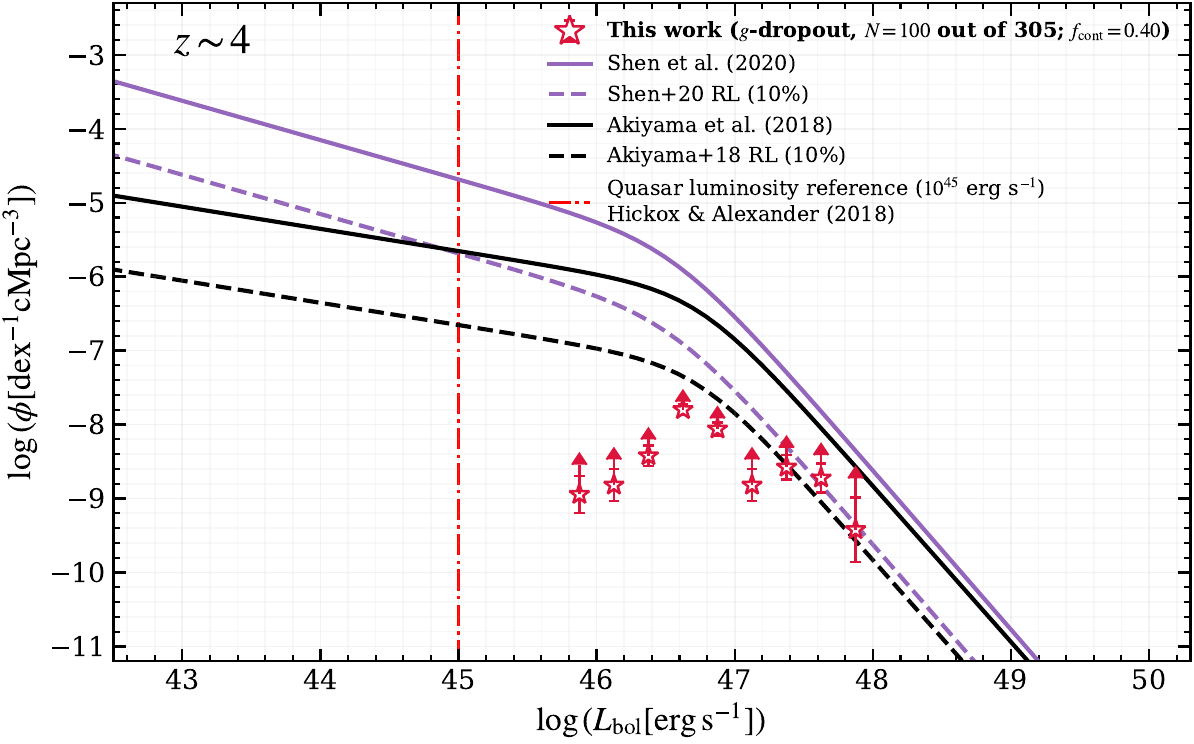}
    \caption{Bolometric luminosity function of the $g$-dropout quasars in this work (open stars), after applying a uniform fiducial contamination fraction of 40\%, compared with the quasar luminosity functions at $z\sim4$ from \citet{shen2020} and \citet{akiyama2018}. The solid curves show the total quasar luminosity functions, while the dashed curves represent the expected radio-loud quasar population assuming a 10\% radio-loud fraction. The vertical red dot-dashed line marks $L_{\rm bol}=10^{45}\,\mathrm{erg\,s^{-1}}$, a commonly used luminosity scale distinguishing quasars from lower-luminosity Seyfert AGNs \citep{hickox2018}.}
    \label{fig:lumi_func_g}
\end{figure*}

\subsection{The quasar luminosity function from this study}
\label{sec:lumi_func}
Previous searches for high-redshift radio AGNs %particularly high-redshift radio galaxies (e.g., \citealt{saxena2019}), are often interpreted within the framework of orientation-based AGN unification,
are often interpreted within the orientation-based AGN unification framework,
in which Type~1 and Type~2 AGNs differ primarily owing to viewing angle and obscuration geometry \citep[e.g.,][]{saxena2019}. 
However, owing to the extreme optical faintness of radio galaxies, even relatively inclusive selection approaches such as adopted here remain limited in their ability to recover comparable numbers of Type~2 systems at high redshift.

%In this study, we use AGN bolometric luminosities derived from SED fitting and bin the sources in luminosity, applying contamination corrections following Equation~\ref{equ:correction} using fractions adapted from \citet{harikane2022}. Owing to the detection limits of the HSC--SSP and the limited availability of multiwavelength photometry at higher redshift, such corrections are only available for $z\sim4$--5. Consequently, only the $g$-dropout sample contains sufficient sources with reliable AGN luminosities, and we therefore derive the quasar bolometric luminosity function only at $z\sim4$.

We use the AGN bolometric luminosities derived from SED
fitting and bin the sources in luminosity. 
%To estimate the impact of foreground interlopers, we apply the magnitude-dependent contamination corrections derived for the HSC Wide-layer dropout samples by \citet{ono2018}. For the magnitude range relevant to our sample, the adopted fractions range from $\sim10$--20\% at $m_{\rm UV}\sim24$--25 and reach $\gtrsim40$--60\% at the brightest magnitudes. These corrections are approximate because they were derived for optically selected HSC dropout samples rather than radio-selected RLAGN candidates.
To estimate the impact of foreground interlopers, we apply the same uniform fiducial contamination fraction of 40\% adopted in Section~\ref{sec:number_density}, assigning a statistical weight of $1-f_{\rm cont}=0.6$ to each source. This correction is approximate because it is based on the contamination indicated by the bright SDSS spectroscopic subsample and is applied uniformly to the full radio-selected RLAGN candidate sample.
Given the available sample size, only the $g$-dropout sample contains sufficient sources with reliable AGN luminosities. We therefore present an exploratory estimate of the bright-end quasar luminosity function at $z\sim4$; uncertainties in contamination, completeness, and redshift determination prevent a robust measurement of the intrinsic luminosity function.

%As shown in Figure~\ref{fig:lumi_func_g}, our measurements are consistent with the luminous end of the quasar luminosity function derived by \citet{akiyama2018} and \citet{shen2020}. The inferred number density is approximately one dex below that of the total quasar population, as expected given that our sample represents the radio-loud subset.
%However, given the substantial uncertainties associated with contamination, completeness, and redshift estimates, this comparison should be regarded primarily as a consistency check rather than a precise determination of the radio-loud quasar luminosity function.}
As shown in Figure~\ref{fig:lumi_func_g}, our contamination-corrected measurements are broadly consistent with the luminous end of the quasar luminosity functions derived by \citet{akiyama2018} and \citet{shen2020}. Given the substantial uncertainties in contamination, completeness, and redshift estimates, this comparison should be regarded as a consistency check rather than a precise determination of the radio-loud quasar luminosity function. The comparison with a nominal 10\% radio-loud fraction is discussed in Section~\ref{sec:RLfraction}.

\subsection{Radio Loud Fraction}\label{sec:RLfraction}

Previous studies based on VLA and SDSS data have established that radio-loud AGNs constitute $\sim10\%$ of the overall quasar population at low to intermediate redshift \citep{kellermann1989, ivezi2002, kellermann2016}.
%Whether this fraction remains unchanged at high redshift is still uncertain.
Whether the radio-loud fraction remains unchanged at high redshift is still under active investigation. Studies of quasars at $z\gtrsim5$ generally find fractions broadly consistent with the canonical $\sim10\%$ value, although the constraints remain limited by sample size, radio sensitivity, and differing definitions of radio loudness \citep{banados2015,gloudemans2021,liu2021,keller2024}. %Dedicated radio follow-up at the highest redshifts likewise indicates that most known quasars are radio quiet, while a minority show radio-loud activity. 
No statistically significant evolution of the radio-loud fraction with redshift has yet been established.

The deep radio and optical data used in this work enable the identification of a substantial population of radio-loud AGN candidates at $z\gtrsim4$. %In contrast, targeted VLA follow-up studies of optically selected $z\gtrsim6$ quasars indicate that most such objects are radio quiet \citep{keller2024}.
At the highest redshifts, dedicated radio follow-up observations of optically selected quasars indicate that the majority of known quasars remain radio quiet, while a minority exhibit radio-loud activity \citep{keller2024}.

%Applying a nominal 10\% scaling to the total quasar luminosity functions of \citet{akiyama2018} and \citet{shen2020} (dashed lines in Figure~\ref{fig:lumi_func_g}), we find that the bright end of our radio-loud sample is broadly consistent with these expectations. This agreement suggests that our selection efficiently recovers the luminous radio-loud quasar population at $z\sim4$. However, the current number-density estimate is derived only from sources with reliable SED fitting, while the total number of dropout candidates is approximately three times larger. Including the full sample would likely increase the inferred number density, indicating that our catalog may represent a more complete census of radio-loud quasars rather than implying a change in the intrinsic radio-loud fraction. At lower luminosities, the observed deficit is likely dominated by optical selection limits and incompleteness in the HSC--SSP, rather than an intrinsic evolution of the radio-loud fraction.

Applying a nominal 10\% scaling to the total quasar luminosity functions of \citet{akiyama2018} and \citet{shen2020} (dashed lines in Figure~\ref{fig:lumi_func_g}), we find that the bright end of our contamination-corrected radio-selected sample is broadly consistent with the expected space density of radio-loud quasars. We emphasize that the 10\% scaling is adopted only as a reference value for comparison and does not assume that the radio-loud fraction is constant with redshift. 
%This comparison is intended only to place our sample in the context of previous studies and should not be interpreted as an independent measurement of the radio-loud fraction.
%Given the uncertainties in the adopted contamination correction, completeness, and redshift estimates discussed above, our data do not provide a robust independent constraint on the radio-loud fraction at $z\gtrsim4$.
%Nevertheless, the observed number densities are broadly consistent with expectations from previous high-redshift studies. 
%Nevertheless, the observed bright-end space densities are broadly consistent with expectations from previous high-redshift studies.
Future spectroscopic follow-up and improved completeness corrections will be required to determine whether the radio-loud fraction evolves significantly toward the highest redshifts.

\section{Conclusion}\label{sec:conclusion}

We constructed a radio AGN catalog by crossmatching compact radio sources from the VLASS Epoch~2 catalog with optical counterparts from the final internal Subaru HSC--SSP Wide-layer release (S23B), and selected high-redshift candidates at $z\gtrsim4$ using the Lyman-break (dropout) technique based on HSC optical colors. Our primary product is a clean dropout-selected candidate catalog consisting of 305 $g$-dropouts, 42 $r$-dropouts, 3 $i$-dropouts, and 2 $z$-dropouts.

We further crossmatched these candidates with ancillary radio surveys, FIRST (1.4\,GHz), TGSS (150\,MHz), and LoTSS DR3 (144\,MHz), to enable multi-frequency radio analysis without imposing any spectral-index pre-selection. 
We also incorporated multiwavelength photometry from VIKING and AllWISE, which enables SED fitting and characterization of AGN properties for the subset with NIR/MIR detections.
The key results of this study are summarized as follows:

\begin{enumerate}
    \item Using dropout selection applied to deep HSC optical identifications of VLASS sources over $\sim1200~\mathrm{deg}^2$, we compile a large sample of candidate radio-loud AGNs at $z\gtrsim4$, extending optimistically to $z\sim7$. %The optical magnitudes cluster at 
    Most candidates have reference-band magnitudes of $i_{\rm AB}\sim24$--26, indicating that most candidates are too faint for systematic spectroscopy in shallower surveys such as SDSS.

    %\item The radio spectral indices of the candidates are predominantly flat to moderately steep ($-1 \lesssim \alpha \lesssim 0$), with a subset of sources showing flatter or positive indices suggestive of spectral curvature or turnover radio SEDs. This demonstrates that conventional ultra-steep-spectrum (USS; $\alpha<-1.3$) selection would miss a substantial fraction of the radio-detected high-$z$ population identified here. Together with the available SDSS spectra and our SED-fitting results, the sample is consistent with being dominated by radio-loud, Type~1 quasar-like AGNs, although targeted spectroscopy is required to quantify the fraction of obscured systems, particularly at the faint end.

    \item The radio spectral indices of the candidates are predominantly flat to moderately steep ($-1\lesssim\alpha\lesssim0$), with a subset showing spectral curvature consistent with turnover or peaked radio spectra. 
    %These results indicate that our dropout-based selection recovers many radio-detected high-redshift candidates that would not satisfy the conventional USS criterion ($\alpha<-1.3$). 
    These results demonstrate that our dropout-selected sample spans a broad range of radio spectral properties without imposing a radio spectral-index criterion.
    The available SDSS spectra indicate that the spectroscopically confirmed high-redshift subset is dominated by radio-loud Type~1 quasars, while the SED-fitting results are consistent with luminous AGN activity. Targeted spectroscopy is required to determine the nature of the full sample.

    \item %For the subset with sufficient NIR/MIR photometry, SED fitting yields bolometric luminosities clustering around $\log(L_{\rm bol}/{\rm erg~s^{-1}})\sim46$--47. %At $z\sim4$, our estimated luminosity function is consistent with tracing the luminous end of the quasar population and lies at approximately the level expected for the radio-loud subset (of order $\sim10\%$) relative to total quasar luminosity functions, within the systematic uncertainties associated with incompleteness and selection effects.
    For the subset with sufficient NIR/MIR photometry, SED fitting yields typical bolometric luminosities of $\log_{10}(L_{\rm AGN}/{\rm erg\,s^{-1}}) \sim46.5$--$47.5$. As an exploratory comparison, the bright end of the $g$-dropout sample is broadly consistent with the luminous end of published quasar luminosity functions and with expectations for the radio-loud subset, although contamination, completeness, and redshift uncertainties prevent a robust luminosity-function determination.
    
    \item %Using a simplified effective-volume approach, we obtain a first-order, luminosity-limited estimate of the dropout-binned comoving number density of radio-loud AGN candidates at $z\sim4$--7. We find a decline toward higher redshift, with a particularly sharp drop from the $r$-dropout ($z\sim5$) to the $i$-dropout ($z\sim6$) regime. While this trend may partly reflect intrinsic evolution of the luminous radio-loud population, it may also be influenced by selection effects and radio detectability limitations, including enhanced inverse-Compton cooling off the cosmic microwave background (``CMB quenching''), which suppresses extended radio emission at high redshift.
    Using a simplified effective-volume approach, we estimate the comoving number density of the dropout-selected RLAGN candidates after applying a uniform fiducial contamination fraction of 40\%. The observed candidate number density declines toward the higher-redshift dropout selections, although this trend may be influenced by contamination, incompleteness, and other selection effects. Robust constraints on the intrinsic evolution of the RLAGN population will require substantially larger spectroscopic samples.

\end{enumerate}

Overall, this work demonstrates that combining deep optical identifications with dropout criteria provides a complementary route to assembling large samples of high-redshift radio AGN candidates without requiring radio spectral-index pre-selection or prior quasar identification.
%Overall, this work demonstrates that deep optical identifications combined with dropout selection provide a complementary, bias-reduced route to assembling large samples of high-$z$ radio AGNs beyond USS selection. 
The resulting catalog provides a wide-area target set for future spectroscopic confirmation and follow-up with optical/NIR facilities and next-generation radio surveys, enabling improved constraints on the demographics and physical properties of radio-loud AGNs approaching the epoch of reionization.

\acknowledgments
This work is supported by the Japan Society for the Promotion of Science (JSPS) KAKENHI (25K01043; K.~Ichikawa).
K.I. also acknowledges support from the JST FOREST Program, Grant Number JPMJFR2466 and the Inamori Research Grants, which helped make this research possible. 
%Z.I. acknowledges the support by the Excellence Cluster ORIGINS which is funded by the Deutsche Forschungsgemeinschaft (DFG, German Research Foundation) under Germany´s Excellence Strategy – EXC-2094 – 390783311.
Y.Z.\ is supported by Japan Society for the Promotion of Science Research Fellowship for Young Scientists.
K.K.\ acknowledges support from JSPS KAKENHI grant Nos JP23K20035 and JP24H00004. 
Y.K.\ acknowledges the support from JST SPRING, Grant Number JPMJSP2108.

The Hyper Suprime-Cam (HSC) collaboration includes the astronomical communities of Japan and Taiwan, and Princeton University.  The HSC instrumentation and software were developed by the National Astronomical Observatory of Japan (NAOJ), the Kavli Institute for the Physics and Mathematics of the Universe (Kavli IPMU), the University of Tokyo, the High Energy Accelerator Research Organization (KEK), the Academia Sinica Institute for Astronomy and Astrophysics in Taiwan (ASIAA), and Princeton University.  Funding was contributed by the FIRST program from the Japanese Cabinet Office, the Ministry of Education, Culture, Sports, Science and Technology (MEXT), the Japan Society for the Promotion of Science (JSPS), Japan Science and Technology Agency  (JST), the Toray Science  Foundation, NAOJ, Kavli IPMU, KEK, ASIAA, and Princeton University.
 
This paper is based on data collected at the Subaru Telescope and retrieved from the HSC data archive system, which is operated by Subaru Telescope and Astronomy Data Center (ADC) at NAOJ. Data analysis was in part carried out with the cooperation of Center for Computational Astrophysics (CfCA) at NAOJ.  We are honored and grateful for the opportunity of observing the Universe from Maunakea, which has the cultural, historical and natural significance in Hawaii.
 
This paper makes use of software developed for Vera C. Rubin Observatory. We thank the Rubin Observatory for making their code available as free software at http://pipelines.lsst.io/. 
 
The Pan-STARRS1 Surveys (PS1) and the PS1 public science archive have been made possible through contributions by the Institute for Astronomy, the University of Hawaii, the Pan-STARRS Project Office, the Max Planck Society and its participating institutes, the Max Planck Institute for Astronomy, Heidelberg, and the Max Planck Institute for Extraterrestrial Physics, Garching, The Johns Hopkins University, Durham University, the University of Edinburgh, the Queen’s University Belfast, the Harvard-Smithsonian Center for Astrophysics, the Las Cumbres Observatory Global Telescope Network Incorporated, the National Central University of Taiwan, the Space Telescope Science Institute, the National Aeronautics and Space Administration under grant No. NNX08AR22G issued through the Planetary Science Division of the NASA Science Mission Directorate, the National Science Foundation grant No. AST-1238877, the University of Maryland, Eotvos Lorand University (ELTE), the Los Alamos National Laboratory, and the Gordon and Betty Moore Foundation.H


% \vspace{5mm}
%\facilities{VLA, Subaru/HSC, VISTA, WISE, LOFAR, SDSS, GMRT}

\clearpage
\appendix
\section{Clean HSC Flags}
%\setlength{\topmargin}{1.0cm} %put to appendix
%\startlongtable
\label{app:hscflag}
\renewcommand{\arraystretch}{1.1}
\begin{deluxetable}{cccc}[h!]
  \tablecaption{Selection criteria for constructing the clean HSC sample\label{tab:hsc_flags}}
  \tablehead{
    \colhead{Parameter} & \colhead{Value} & \colhead{Band} & \colhead{Comment}
  }
  \startdata
    \texttt{isprimary}         & True  & \dots        & Object is a primary one with no deblended children \\
    \texttt{pixelflags\_edge}  & False & \textit{grizy} & Located within image boundaries \\
    \texttt{pixelflags\_saturatedcenter} & False & \textit{grizy} & None of the central 3 $\times$ 3 pixels of an object is saturated \\
    \texttt{pixelflags\_bad} & False & \textit{grizy} & None of the pixels in the footprint of an object is labeled as bad \\
    \texttt{mask\_brightstar\_ghost} & False & \textit{grizy} & None of the pixels in the footprint of an object is close to the ghost masks \\
    \texttt{mask\_brightstar\_blooming} & False & \textit{grizy} & None of the pixels in the footprint of an object is close to the bloom masks \\
    \texttt{mask\_brightstar\_channel\_stop} & False & \textit{y} & None of the pixels in the footprint of an object is close to the channel stop masks. \\
    \texttt{inputcount\_value} & $\geq3 / \geq5$ & \textit{gr/izy} & The number of exposures is equal to or larger than 3/5 in \textit{gr/izy} \\
    \texttt{sdsscentroid\_flag} & False & \textit{ri} for $g$-dropout & Object centroid measurement has no problem \\
      & False & \textit{iz} for $r$-dropout & \dots \\
      & False & \textit{zy} for $i$-dropout & \dots \\
      & False & \textit{y} for $z$-dropout & \dots \\
    \texttt{cmodel\_flag} & False & \textit{gri} for $g$-dropout & Cmodel flux measurement has no problem \\
      & False & \textit{riz} for $r$-dropout & \dots \\
      & False & \textit{izy} for $i$-dropout & \dots \\
      & False & \textit{zy} for $z$-dropout & \dots \\
    \texttt{merge\_peak} & True & \textit{ri} for $g$-dropout & Detected in $r$ and $i$.\\
      & False/True & \textit{g/iz} for $r$-dropout & Undetected in $g$ and detected in $i$ and $z$.\\
       & False/True & \textit{gr/zy} for $i$-dropout & Undetected in $g$ and $r$, and detected in $z$ and $y$.\\
      & False/True & \textit{gri/y} for $z$-dropout & Undetected in $g$, $r$, and $i$, and detected in $y$.\\
    \texttt{blendedness\_abs} & $<0.2$ & \textit{ri} for $g$-dropout & The target photometry is not significantly affected by neighbors.\\
      & $<0.2$ & \textit{iz} for $r$-dropout & \dots\\
    & $<0.2$ & \textit{zy} for $i$-dropout & \dots\\
        & $<0.2$ & \textit{y} for $z$-dropout & \dots\\
        \texttt{convolvedflux\_0\_20\_flag} & False & \textit{gri} for $g$-dropout & The flux measurement within the $2''$ diameter aperture has no problem \\
         & False & \textit{riz} for $r$-dropout & \dots \\
         & False & \textit{izy} for $i$-dropout & \dots \\
         & False & \textit{zy} for $z$-dropout & \dots \\
\enddata
\end{deluxetable}

\section{Examples of Rejected $z$-Dropout Candidates in Warp Images}
\label{app:badz}
%In this section, we present representative examples of spurious $z$-dropout candidates that were removed during the visual inspection stage of the cleaning process. These cases illustrate typical failure modes in the HSC coadded images, including false detections arising from transient or moving objects in individual warp exposures and incorrect optical--radio associations. 
In this section, we present representative examples of spurious $z$-dropout candidates removed during the visual inspection stage. These cases illustrate typical failure modes in the HSC coadded images, including false detections caused by transient or moving objects in individual warp exposures and incorrect optical--radio associations. Each numbered stamp shows an individual valid warp exposure, and the number in the upper-left corner identifies the corresponding valid-frame index. %A $2''$ scale bar is shown in each stamp.
A $2''$ scale bar is shown in each panel.

%\begin{figure}[h!]
 %   \centering
  %  \includegraphics[width=\linewidth]{tract_8763_same_contrast.pdf}
   % \caption{Example of a misassociated optical--radio match caused by a detection present in only a single warp image. The apparent source in the coadd is produced by a transient feature in an individual exposure, leading to an erroneous optical counterpart.}
  %  \label{fig:z_warp_1}
%\end{figure}
%\begin{figure}[h!]
  %  \centering
 %   \includegraphics[width=0.45\linewidth]{tract_9243_same_contrast.pdf}
  %  \caption{False $y$-band detection produced by a bright moving object in one of the warp images, which is subsequently smoothed into the coadded image and mistakenly identified as a real source.}
  %  \label{fig:z_warp_2}
%\end{figure}
%\begin{figure}[h!]
   % \centering
   % \includegraphics[width=0.7\linewidth]{tract_10061_same_contrast.pdf}
   % \caption{Example of a spurious $z$-dropout candidate resulting from both an incorrect optical--radio association and contamination by a moving object in individual warp exposures, leading to a false coadd detection.}
    %\label{fig:z_warp_3}
%\end{figure}

\begin{figure}[h!]
    \centering
    \includegraphics[width=\linewidth]{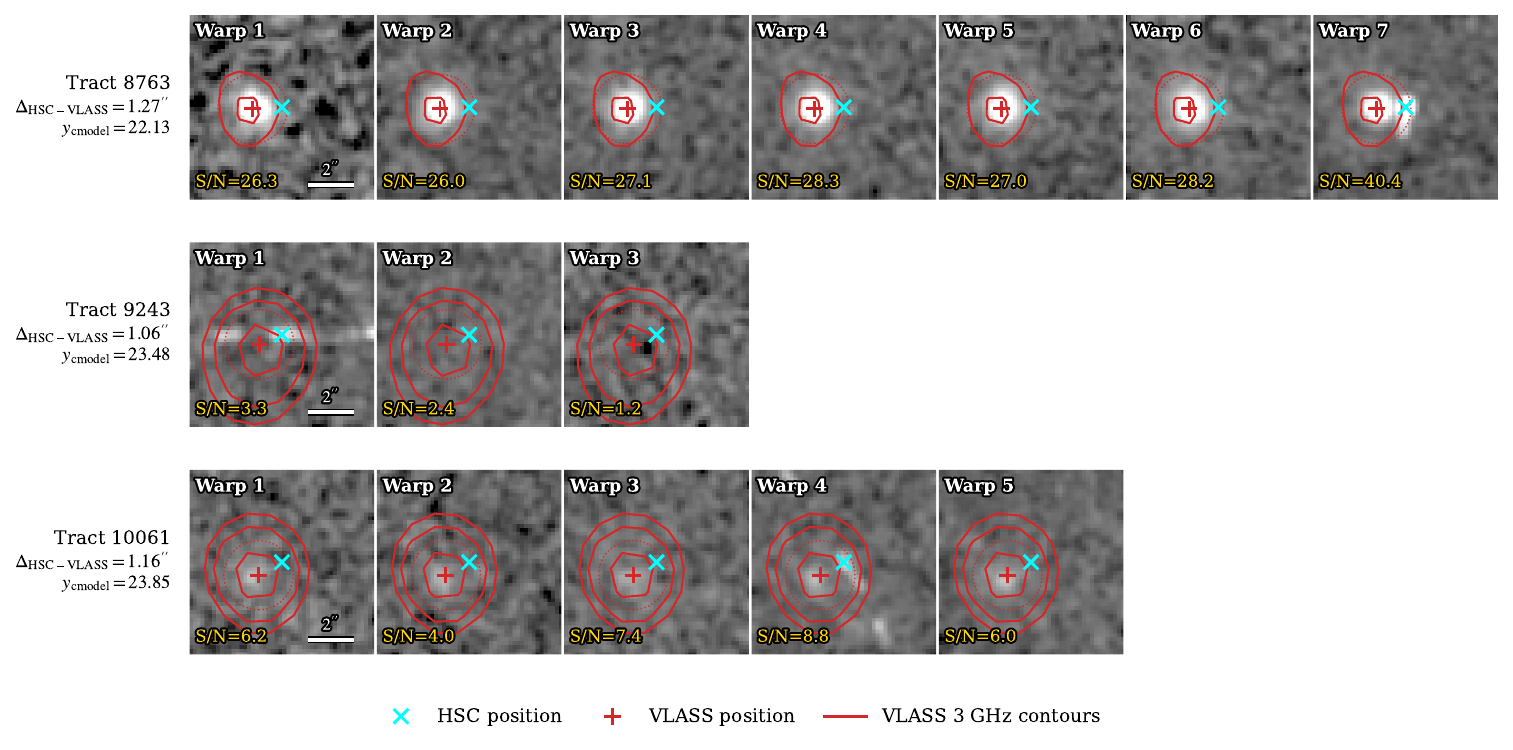}
    \caption{Individual HSC $y$-band warp images of the three rejected $z$-dropout candidates, with VLASS 3~GHz radio contours overlaid to examine the optical--radio associations. Each row corresponds to one candidate, and the panels from left to right show the individual valid warp exposures contributing at the source position; ``Warp 1'', ``Warp 2'', etc. indicate the corresponding valid-frame indices. Cyan crosses mark the HSC optical positions, red plus symbols mark the VLASS positions, and solid red contours show the VLASS 3~GHz emission at $3$, $5$, $10$, $20$, and $40\sigma$. A $2^{\prime\prime}$ scale bar is shown in the first panel of each row, and the yellow labels give the approximate aperture S/N in each warp. From top to bottom, the examples show a misassociated optical--radio match associated with a transient feature in an individual exposure, a false $y$-band detection produced by a moving object, and a spurious $z$-dropout candidate resulting from both an incorrect optical--radio association and moving-object contamination.}
    \label{fig:z_warps}
\end{figure}

\section{List of SDSS Identified Sources}
\label{app:sdssz}

%In this section, we present the list of sources that is detectable by SDSS (i.e. $i_{\rm mag}\leq21.5$), obtained a valid match for SDSS spectra, and visually checked for line identification. 17 of the source have their spectroscopic redshift consistent with their $g$-dropout colors (Figure~\ref{fig:sdss_spec}) and the rest 12 low-$z$ contaminant with line emission more than 3-$\sigma$ for redshift determination. 
In this section, we list the 29 $g$-dropout candidates with usable SDSS spectra ($\mathrm{S/N}\gtrsim3$) that were visually inspected for line identification. 17 sources have spectroscopic redshifts consistent with the expected $g$-dropout redshift range (Figure~\ref{fig:sdss_spec}), while the remaining 12 are low-redshift contaminants with spectral features sufficient for reliable redshift
determination.

% list of sources with spec-z (Table) SDSS + FOCAS(?) full table + 3-4 representative targets.
\vspace{-2em}
\begin{longtable}{llllll}
    \label{tab:sdss_source}\\
    \caption{Sources with SDSS spectroscopy}\\
    \hline
    ID & RA (J2000) & Dec (J2000) &  $z_{\rm spec}$ & $i_{\rm mag}$ & Identification line emission\\
    \hline
\endfirsthead
\hline
 ID & RA (J2000) & Dec (J2000) &  $z_{\rm spec}$ & $i_{\rm mag}$ &
Identified emission lines\\
\hline
\endhead
\hline
\endfoot
\hline
\endlastfoot
    HSC J223535+003602 & 22:35:35.591 & +00:36:02.01 & 3.8725 & 20.3032 & Ly$\alpha$, C\,\textsc{iv}\\
HSC J231448+020151 & 23:14:48.717 & +02:01:51.07 & 4.1141 & 19.7401 & Ly$\alpha$, C\,\textsc{iv}\\
HSC J233432+023321 & 23:34:32.023 & +02:33:21.21 & 4.1018 & 20.0345 &  Ly$\alpha$, C\,\textsc{iv}\\
HSC J153830+424405 & 15:38:30.718 & +42:44:05.68 & 4.1118 & 20.7969 & Ly$\alpha$, C\,\textsc{iv}\\
HSC J144855+434905 & 14:48:55.200 & +43:49:05.15 & 0.3617 & 19.6341 & O\,\textsc{[iii]}, H$\alpha$\\
HSC J150544+433824 & 15:05:44.603 & +43:38:24.69 & 4.6800 & 19.1168 & Ly$\alpha$, C\,\textsc{iv}\\
HSC J020630-032847 & 02:06:30.483 & -03:28:47.17 & 3.5254 & 20.6529 & Ly$\alpha$, C\,\textsc{iv}\\
HSC J022320-031824 & 02:23:20.704 & -03:18:24.27 & 3.8651 & 19.3650 & Ly$\alpha$, C\,\textsc{iv}\\
HSC J141315-014220 & 14:13:15.304 & -01:42:20.95 & 0.3803 & 19.5935 &O\,\textsc{[ii]},O\,\textsc{[iii]}, H$\alpha$\\
HSC J000252-000330 & 00:02:52.726 & -00:03:30.99 & 3.6904 & 20.9594 & Ly$\alpha$, C\,\textsc{iv}\\
HSC J105407-012342 & 10:54:07.139 & -01:23:42.34 & 3.9845 & 20.9389 & Ly$\alpha$, C\,\textsc{iv}\\
HSC J105320-001649 & 10:53:20.426 & -00:16:49.65 & 4.2969 & 19.0413 & Ly$\alpha$, C\,\textsc{iv}\\
HSC J123503-000331 & 12:35:03.047 & -00:03:31.75 & 4.7023 & 20.4025 & Ly$\alpha$, C\,\textsc{iv}\\
HSC J011341+010608 & 01:13:41.112 & +01:06:08.63 & 0.2809 & 18.4618 &O\,\textsc{[ii]},O\,\textsc{[iii]}, H$\alpha$\\
HSC J092356+012001 & 09:23:56.458 & +01:20:02.00 & 0.3797 & 19.6403 &O\,\textsc{[ii]},O\,\textsc{[iii]}, H$\alpha$\\
HSC J092004+012115 & 09:20:04.529 & +01:21:15.33 & 3.9026 & 20.6726 & Ly$\alpha$, C\,\textsc{iv}\\
HSC J093502+004916 & 09:35:02.996 & +00:49:16.62 & 0.3480 & 17.8864 &H$\alpha$\\
HSC J094822+005554 & 09:48:22.971 & +00:55:54.42 & 3.8801 & 20.0944 & Ly$\alpha$, C\,\textsc{iv}\\
HSC J130002+011823 & 13:00:02.168 & +01:18:23.14 & 4.6124 & 19.1448 & Ly$\alpha$, C\,\textsc{iv}\\
HSC J143634+005111 & 14:36:34.495 & +00:51:11.87 & 3.6858 & 21.2429 & Ly$\alpha$, C\,\textsc{iv}\\
HSC J113437+025343 & 11:34:37.059 & +02:53:43.09 & 0.3222 & 19.3960 &O\,\textsc{[ii]},O\,\textsc{[iii]}, H$\alpha$, H$\beta$\\
HSC J114054+023024 & 11:40:54.791 & +02:30:24.63 & 0.4536 & 18.9249 &O\,\textsc{[ii]},O\,\textsc{[iii]}\\
HSC J002504+030615 & 00:25:04.918 & +03:06:15.70 & 0.6033 & 19.2995 &O\,\textsc{[ii]},O\,\textsc{[iii]}\\
HSC J005158+042740 & 00:51:58.994 & +04:27:40.13 & 0.2637 & 18.5058 &O\,\textsc{[ii]},O\,\textsc{[iii]}, H$\alpha$, H$\beta$\\
HSC J012126+034706 & 01:21:26.147 & +03:47:06.76 & 4.1333 & 18.9370 & Ly$\alpha$, C\,\textsc{iv}\\
HSC J112509+033955 & 11:25:09.574 & +03:39:55.06 & 0.2614 & 17.7110 &O\,\textsc{[ii]},O\,\textsc{[iii]}, H$\alpha$\\
HSC J005405+044513 & 00:54:05.479 & +04:45:13.43 & 4.2655 & 21.4651 & Ly$\alpha$, C\,\textsc{iv}\\
HSC J135637+430403 & 13:56:37.047 & +43:04:03.77 & 0.1932 & 18.3133 &O\,\textsc{[ii]},O\,\textsc{[iii]}, H$\alpha$, H$\beta$\\
HSC J141647+425112 & 14:16:47.711 & +42:51:12.28 & 0.3478 & 18.6316 &O\,\textsc{[ii]},O\,\textsc{[iii]}, H$\alpha$, H$\beta$\\
\hline
\end{longtable}
% Note: based on visual checking

\section{$k$-Correction for Radio Loudness}
\label{app:k_corr}
\begin{figure}[h!]
    \centering
    \includegraphics[width=0.5\linewidth]
    {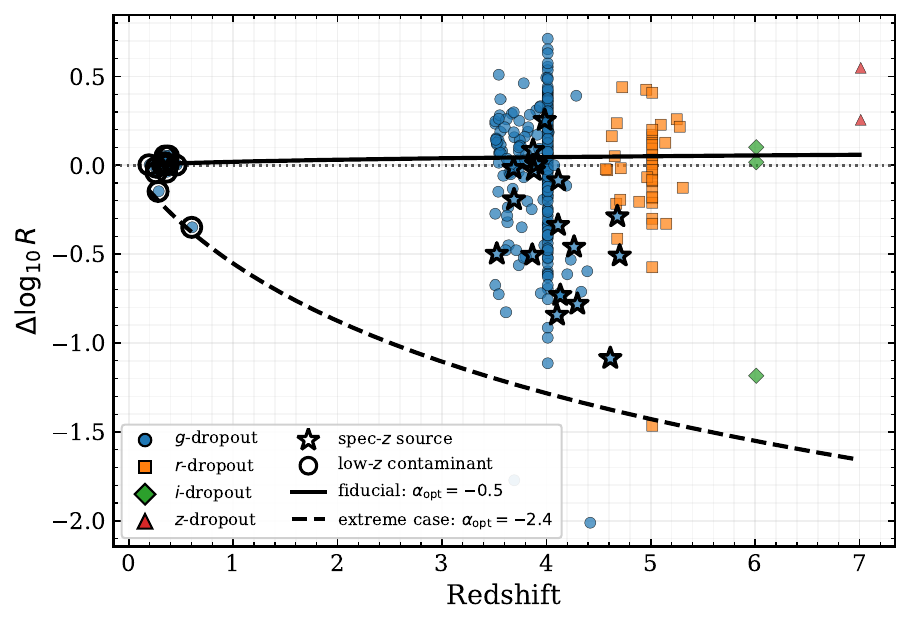}
    \caption{Effect of the adopted first-order
    $k$-correction on the radio loudness. Colored symbols
    denote the $g$-, $r$-, $i$-, and $z$-dropout samples and
    show the logarithmic correction $\Delta\log_{10}R$ as a function of redshift. Open black stars mark spectroscopically confirmed high-redshift sources, while open black circles mark spectroscopically confirmed low-redshift contaminants. The solid and dashed black curves show the corrections for $\alpha_{\rm opt}=-0.5$ and $-2.4$, respectively. Gray vertical lines indicate the range between these two assumptions for individual sources.}
    \label{fig:radio_loudness_kcorr}
\end{figure}
Following the first-order treatment of \citet{ich21}, and assuming power-law radio and optical spectra, $f_{\nu}\propto\nu^{\alpha}$, the logarithmic $k$-correction applied to the radio loudness in Equation~\eqref{equ:RKcorr} is given by

\begin{equation}
\Delta\log_{10}R
=
\left(
\alpha_{\rm opt}
-
\alpha_{\rm radio}
\right)
\log_{10}(1+z),
\label{equ:delta_Rkcorr}
\end{equation}
%where $R_{\rm obs}$ is defined in Section~\ref{sec:radio_loudness}, $\alpha_{\rm radio}$ is the representative radio spectral index adopted in Section~\ref{sec:radio_color}, and $\alpha_{\rm opt}$ is the optical continuum spectral index. We adopt $\alpha_{\rm opt}=-0.5$ as the fiducial value \citep{ich21} and examine $\alpha_{\rm opt}=-2.4$ as a steeper illustrative case.}
where $\alpha_{\rm radio}$ is the representative radio
spectral index adopted in Section~\ref{sec:radio_color}, and
$\alpha_{\rm opt}$ is the optical continuum spectral index.
We adopt $\alpha_{\rm opt}=-0.5$ as the fiducial value
\citep{ich21} and examine $\alpha_{\rm opt}=-2.4$ as a
steeper illustrative case.
Figure~\ref{fig:radio_loudness_kcorr} shows that the correction becomes increasingly redshift-dependent, with its sign and amplitude determined by the adopted optical and radio spectral indices. Under the fiducial assumption, the overall sample remains in the high-$R_{\rm kcorr}$ regime, while the steeper optical slope illustrates the associated systematic uncertainty.

\section{CIGALE Configuration}
\label{app:cigale}
\begin{table}[h!]
    \caption{Model parameters adopted for the CIGALE fits}
    \centering
    \begin{tabular}{ccc}
    \hline
       Module  & Parameter & Value \\
       \hline
       \hline
        \texttt{Skirtor2016} & t & 7, 9, 11 \\
         & pl & 0.0, 0.5, 1.0 \\
        & q & 0.0, 0.5, 1.0 \\
        & oa & 30, 40, 50, 60 \\
        & R & 10, 20, 30 \\
        & i & 0, 10, 20 \\
        & delta & -0.25, 0 \\
        & fracAGN & 0.99 \\
        & law & 0, 1 \\
        & EBV & 0.03, 0.1, 0.3, 0.5, 1.0, 2.0 \\
        & temperature & 50.0, 75.0, 100.0, 150.0, 200.0, 300.0 \\
        & emissivity & 1.0, 1.6, 2.0 \\
    \hline
    \texttt{Redshifting} & $g$-dropout & 3.5, 3.6, 3.7, 3.8, 3.9, 4.0, 4.1, 4.2, 4.3, 4.4, 4.5 \\
    &$r$-dropout &4.5, 4.6, 4.7, 4.8, 4.9, 5.0, 5.1, 5.2, 5.3, 5.4, 5.5, 5.6, 5.7, 5.8\\
    \hline
    \hline
    
    \end{tabular}
    \label{tab:cigale_config}
\end{table}

For the CIGALE SED fitting, we model only the AGN component using the SKIRTOR torus models \citep{stalevski2012,stalevski2016} together with the \texttt{redshifting} module, which also includes intergalactic medium absorption. No stellar or host-galaxy components are included, as our sources are strongly AGN dominated in both the optical and infrared. Photometric redshifts are obtained by fitting over predefined redshift grids appropriate for each dropout class. 
Table~\ref{tab:cigale_config} summarizes the adopted parameter space.

\section{Possible Contamination from Type~2 AGN Hosts}
\label{app:type 2}
To assess the potential presence of Type~2 AGN hosts within our sample, we perform a blind SED-fitting analysis for sources without spectroscopic redshift confirmation by force-fitting models that include a Type~2 AGN component and host-galaxy emission. These fits systematically result in unrealistically high stellar mass estimates, indicating that the observed emission is dominated by the AGN rather than by the stellar host.

Based on this result, we conclude that contamination from Type~2 AGN hosts is negligible in our sample. We therefore treat the sources as radio-loud quasars and adopt a pure AGN model for the subsequent SED fitting analysis.

\section{Low-z contaminants}
\label{app:low-z}
\begin{figure}[h!]
    \centering
    \includegraphics[width=0.6\linewidth]{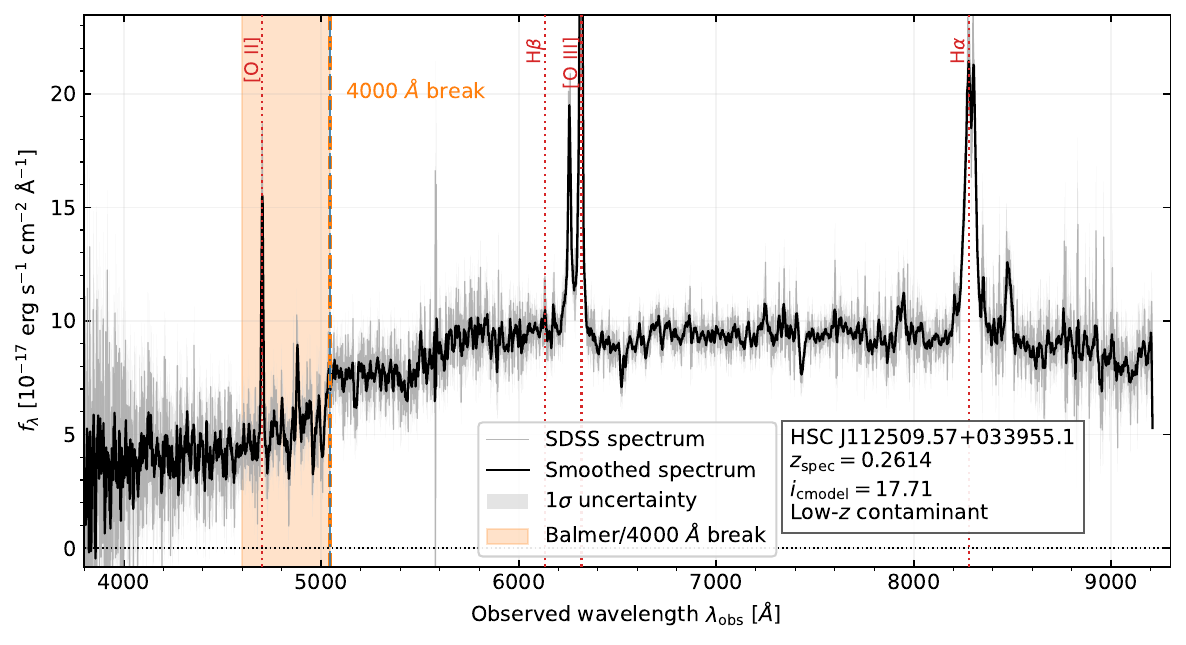}
    \caption{  Optical spectrum of a spectroscopically confirmed low-redshift contaminant in the dropout-selected sample. The spectrum exhibits a prominent Balmer/4000\,\AA\ break and strong [O\,\textsc{iii}] emission lines.}
    \label{fig:low_z}
\end{figure}
Among the 29 spec-$z$ available $g$-dropout sources from the SDSS survey, 12 objects in our sample are identified as low-$z$ AGN interlopers. All of these low-$z$ interlopers are found at $z\sim0.5$, since they exhibit galaxy-dominated spectra with prominent Balmer/4000\,\AA\ breaks, together with strong [O\,\textsc{iii}] emission lines arising from the AGN narrow-line region. These spectral features mimic the Lyman break used in the color selection, allowing them to satisfy the dropout criteria despite their much lower redshifts, as illustrated in the left panel of Figure~\ref{fig:color-color}.

\begin{figure}[h]
    \centering
    \includegraphics[width=0.7\linewidth]{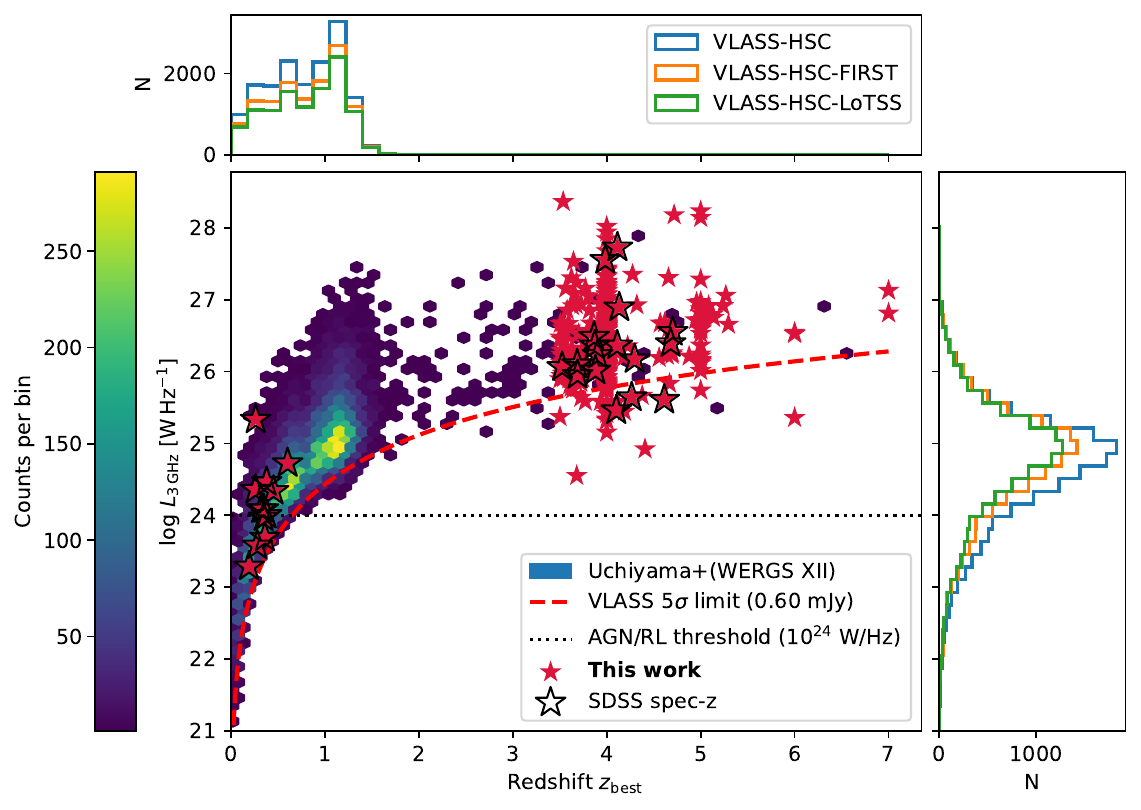}
    %\caption{Comparison of our dropout-selected sources (red stars) with the full HSC--VLASS radio AGN population (density map) and the VLASS $5\sigma$ sensitivity limit (Uchiyama et al.\ in prep.), together with the radio-loud AGN threshold.}
    \caption{Rest-frame 3\,GHz radio luminosity versus redshift, as in Figure~\ref{fig:uchiyama_overlay}, with the spectroscopically confirmed low-redshift contaminants overplotted. The background density map shows the parent HSC--VLASS radio-source population from \citet{uch26}. The contaminants occupy the broad radio-luminosity range of the parent population at their spectroscopic redshifts.}
    \label{fig:uchiyama_overlay_with_low_z}
\end{figure}

Figure~\ref{fig:low_z} shows a representative spectrum, while the remaining 11 interlopers exhibit broadly similar break-dominated spectral features. Figure~\ref{fig:uchiyama_overlay_with_low_z} compares their radio luminosities and spectroscopic redshifts with those of the confirmed high-redshift sources and the broader HSC--VLASS radio-source population. At their spectroscopic redshifts, the contaminants occupy the broad radio-luminosity range of the parent population, indicating that their low-redshift nature is identified primarily from their optical spectra rather than from their radio properties.

%For candidates without spectroscopic confirmation, the nature of the contamination remains uncertain. Based on the evolutionary tracks shown in Figure~\ref{fig:color-color}, some contaminants may be dust-obscured galaxies at intermediate redshift ($z\sim1$--3), whose reddened continua produce colours similar to those of high-redshift dropout-selected sources. Spectroscopic follow-up observations will be required to quantify the contamination fraction and determine the physical nature of these objects.}

\section{Catalog Columns}
\label{app:catalog_column}
We present the columns provided in the HSC-VLASS dropout-selected radio AGN catalog in Table~\ref{tab: catalog_columns}. The catalog includes radio properties from VLASS, which serves as the reference survey at 3~GHz, as well as multiwavelength data from HSC (optical to near-infrared; \textit{grizy}), VIKING (near-infrared; \textit{ZYJHKs}), WISE (mid-infrared; \textit{W1, W2, W3, W4}), FIRST (1.4~GHz), LoTSS (144~MHz), and TGSS (150~MHz). In addition, the catalog contains derived quantities calculated in this work.
\vspace{-2em}
\begin{longtable}{llll}
\label{tab: catalog_columns}\\
\caption{Columns for the Main Catalog} \\
\hline
Column name & Type &  Units &  Description \\
\hline
\endfirsthead
\hline
Column name &  Type &  Units &  Description \\
\hline
\endhead
\hline
\endfoot
\hline
\endlastfoot

\hline
\hline
\multicolumn{4}{c}{Columns for VLASS \citep{gordon2021}} \\

Component\_name & String & \dots & Unique name of the PyBDSF component\\
Component\_id & Integer & \dots & Component identifier\\
Isl\_id & Integer & \dots & Island identifier\\

RA\_vlass & Double & deg & Right ascension of component\\
DEC\_vlass & Double & deg & Declination of component\\
E\_RA\_vlass & Double & deg & Error in right ascension\\
E\_DEC\_vlass & Double & deg & Error in declination\\

Total\_flux\_vlass & Double & mJy & Integrated component flux density\\
E\_Total\_flux\_vlass & Double & mJy & Error in integrated flux density\\
Peak\_flux\_vlass & Double & mJy beam$^{-1}$ & Peak component flux density\\
E\_Peak\_flux\_vlass & Double & mJy beam$^{-1}$ & Error in peak flux density\\

Maj\_vlass & Double & arcsec & Major axis (FWHM)\\
E\_Maj\_vlass & Double & arcsec & Error in major axis\\
Min\_vlass & Double & arcsec & Minor axis (FWHM)\\
E\_Min\_vlass & Double & arcsec & Error in minor axis\\
PA\_vlass & Double & deg & Position angle east of north\\
E\_PA\_vlass & Double & deg & Error in position angle\\

Isl\_Total\_flux & Double & mJy & Integrated island flux density\\
E\_Isl\_Total\_flux & Double & mJy & Error in island integrated flux\\
Isl\_rms & Double & mJy beam$^{-1}$ & Island rms\\
Isl\_mean & Double & mJy beam$^{-1}$ & Island mean\\
Resid\_Isl\_rms & Double & mJy beam$^{-1}$ & Residual island rms\\
Resid\_Isl\_mean & Double & mJy beam$^{-1}$ & Residual island mean\\

RA\_max & Double & deg & Right ascension of peak pixel\\
DEC\_max & Double & deg & Declination of peak pixel\\
E\_RA\_max & Double & deg & Error in peak-pixel right ascension\\
E\_DEC\_max & Double & deg & Error in peak-pixel declination\\

S\_Code & String & \dots & PyBDSF source type\\

DC\_Maj & Double & arcsec & Deconvolved major axis (FWHM)\\
E\_DC\_Maj & Double & arcsec & Error in deconvolved major axis\\
DC\_Min & Double & arcsec & Deconvolved minor axis (FWHM)\\
E\_DC\_Min & Double & arcsec & Error in deconvolved minor axis\\
DC\_PA & Double & deg & Deconvolved position angle\\
E\_DC\_PA & Double & deg & Error in deconvolved position angle\\

Maj\_img\_plane & Double & arcsec & Major axis in image plane\\
E\_Maj\_img\_plane & Double & arcsec & Error in image-plane major axis\\
Min\_img\_plane & Double & arcsec & Minor axis in image plane\\
E\_Min\_img\_plane & Double & arcsec & Error in image-plane minor axis\\
PA\_img\_plane & Double & deg & Image-plane position angle\\
E\_PA\_img\_plane & Double & deg & Error in image-plane position angle\\

DC\_Maj\_img\_plane & Double & arcsec & Deconvolved image-plane major axis\\
E\_DC\_Maj\_img\_plane & Double & arcsec & Error in deconvolved image-plane major axis\\
DC\_Min\_img\_plane & Double & arcsec & Deconvolved image-plane minor axis\\
E\_DC\_Min\_img\_plane & Double & arcsec & Error in deconvolved image-plane minor axis\\
DC\_PA\_img\_plane & Double & deg & Deconvolved image-plane position angle\\
E\_DC\_PA\_img\_plane & Double & deg & Error in deconvolved image-plane position angle\\

Tile & String & \dots & VLASS observing tile\\
Subtile & String & \dots & VLASS observing subtile\\

QL\_image\_RA & Double & deg & Quick-look image center right ascension\\
QL\_image\_DEC & Double & deg & Quick-look image center declination\\

NVSS\_distance & Double & arcsec & Separation to NVSS counterpart\\
FIRST\_distance & Double & arcsec & Separation to FIRST counterpart\\

Peak\_to\_ring & Double & \dots & Ratio of peak flux to maximum flux within an annulus of $5\arcsec<r<10\arcsec$\\
Duplicate\_flag & Integer & \dots & Duplicate flag\\
Quality\_flag & Integer & \dots & Quality flag\\

Source\_name\_vlass & String & \dots & Parent radio-source identifier\\
Source\_type & String & \dots & Source morphology classification\\

NN\_dist & Double & arcsec & Distance to nearest VLASS component\\

BMAJ & Double & arcsec & Beam major axis\\
BMIN & Double & arcsec & Beam minor axis\\
BPA & Double & deg & Beam position angle\\

Best\_neuron\_y & Integer & pixel & PyBDSF neuron $y$ coordinate\\
Best\_neuron\_x & Integer & pixel & PyBDSF neuron $x$ coordinate\\
Neuron\_dist & Double & pixel & Distance to nearest neuron\\
P\_sidelobe & Double & \dots & Probability of being a sidelobe\\
\hline
\multicolumn{4}{c}{Columns for HSC S23B Wide \citep{aihara2022}} \\
ra & Double & deg & \dots\\
dec & Double & deg & \dots \\
object\_id & Integer & \dots & unique id for each HSC object\\
g/r/i/z/y\_cmodel\_mag & Float & mag & total g/r/i/z/y-band flux\\
g/r/i/z/y\_cmodel\_magerr & Float & mag & total g/r/i/z/y-band flux error\\
g/r/i/z/y\_cmodel\_flux & Float & nano-Jy & total g/r/i/z/y-band flux\\
g/r/i/z/y\_cmodel\_fluxerr & Float & nano-Jy & total g/r/i/z/y-band flux error\\
\hline
\multicolumn{4}{c}{Columns for FIRST \citep{helfand2015}} \\

RA\_first & String & \dots & Right Ascension J2000 \\
DEC\_first & String & \dots & Declination J2000 \\
SIDEPROB & Float & \dots & [0,1] Probability of being a side lobe \\
FPEAK & Double & mJy beam$^{-1}$ & Peak flux density at 1.4~GHz \\
FINT & Double & mJy & Integrated flux density at 1.4~GHz \\
RMS & Float & mJy & Local noise estimate \\
MAJOR & Float & arcsec & Major axis (FWHM) \\
MINOR & Float & arcsec & Minor axis (FWHM) \\
POSANG & Float & deg & [0/180] Position angle \\

FITTED\_MAJOR & Float & arcsec & Fitted major axis (FWHM) \\
FITTED\_MINOR & Float & arcsec & Fitted minor axis (FWHM) \\
FITTED\_POSANG & Float & deg & Fitted position angle \\

FLDNAME & String & \dots & FIRST observing field \\

NSDSS & String & \dots & [0/10] Number of SDSS-DR10 counterparts \\
SDSS\_SEP & Float & arcsec & Separation to nearest SDSS counterpart \\
SDSS\_MAG & Float & mag & Magnitude of nearest SDSS counterpart \\
SDSS\_CLASS & Character & \dots & [sg-] SDSS class: s=star, g=galaxy \\

NTMASS & String & \dots & Number of 2MASS counterparts \\
TMASS\_SEP & Float & arcsec & Separation to nearest 2MASS counterpart \\
TMASS\_MAG & Float & mag & Magnitude of nearest 2MASS counterpart \\

YEAR & Integer & yr & Observation year \\
MJD & Double & day & Mean Modified Julian Date \\
MJDRMS & Double & day & RMS of Modified Julian Date \\
MJDSTART & Double & day & Observation start (MJD) \\
MJDSTOP & Double & day & Observation end (MJD) \\

\hline
\multicolumn{4}{c}{Columns for LoTSS \citep{shimwell2026}} \\
Source\_Name\_lotss & String & \dots & The radio name of the source, automatically generated from RA and DEC \\
RA\_lotss & Double & deg & Right ascension (J2000) (RA) \\
E\_RA\_lotss & Double & deg & Uncertainty in right ascension \\
DEC\_lotss & Double & deg & Declination (J2000) (DEC) \\
E\_DEC\_lotss & Double & deg & Uncertainty in declination \\
Peak\_flux\_lotss & Double & mJy beam$^{-1}$ &
The peak Stokes I flux density per beam of the source (Peak\_flux) \\
E\_Peak\_flux\_lotss & Double & mJy beam$^{-1}$ &
Uncertainty in peak flux density \\
Total\_flux\_lotss & Double & mJy &
The total, integrated Stokes I flux density of the source \\
& & & at the reference frequency (Total\_flux) \\
E\_Total\_flux\_lotss & Double & mJy &
Uncertainty in integrated flux density \\
Maj\_lotss & Float & arcsec &
FWHM of the major axis of the source, INCLUDING convolution \\
& & & with the 6-arcsec LOFAR beam (Maj) \\
E\_Maj\_lotss & Float & arcsec & Uncertainty in Maj \\
Min\_lotss & Float & arcsec &
FWHM of the minor axis of the source, INCLUDING convolution \\
& & & with the 6-arcsec LOFAR beam (Min) \\
E\_Min\_lotss & Float & arcsec & Uncertainty in Min \\
DC\_Maj\_lotss & Float & arcsec &
The FWHM of the major axis of the source, after de-convolution \\
& & & with the 6-arcsec LOFAR beam (DC\_Maj) \\
E\_DC\_Maj\_lotss & Float & arcsec & Uncertainty in DC\_Maj \\
DC\_Min\_lotss & Float & arcsec &
The FWHM of the minor axis of the source, after de-convolution \\
& & & with the 6-arcsec LOFAR beam (DC\_Min) \\
E\_DC\_Min\_lotss & Float & arcsec & Uncertainty in DC\_Min \\
PA\_lotss & Float & deg &
The position angle of the major axis of the source measured east of north, \\
& & & after de-convolution with the 6-arcsec LOFAR beam (PA) \\
E\_PA\_lotss & Float & deg & Uncertainty in PA \\
DC\_PA\_lotss & Float & deg &
The position angle of the major axis of the source measured east of north, \\
& & & after de-convolution with the 6-arcsec LOFAR beam (DC\_PA) \\
E\_DC\_PA\_lotss & Float & deg & Uncertainty in DC\_PA \\
S\_Code\_lotss & String & \dots &
A code that defines the source structure \\
& & & in terms of the fitted Gaussian components \\
Isl\_rms\_lotss & Float & mJy beam$^{-1}$ &
Local island rms noise \\
Mosaic\_ID\_lotss & String & \dots &
Identifier of the LoTSS mosaic \\
Number\_Pointings\_lotss & Integer & \dots &
Number of LOFAR pointings contributing to the mosaic \\
Masked\_Fraction\_lotss & Float & \dots &
The fraction of the source that is in the CLEAN mask (Masked\_Fraction) \\

\hline
\multicolumn{4}{c}{Columns for TGSS ADR1 \citep{intema2017}} \\

Source\_name\_tgss & String & \dots & Source designation \\

RA\_tgss & Double & deg & Right ascension (J2000) \\
E\_RA\_tgss & Double & deg & Uncertainty in right ascension \\

DEC\_tgss & Double & deg & Declination (J2000) \\
E\_DEC\_tgss & Double & deg & Uncertainty in declination \\

Total\_flux\_tgss & Double & mJy &
Integrated flux density at 150 MHz \\
E\_Total\_flux\_tgss & Double & mJy &
Uncertainty in integrated flux density \\

Peak\_flux\_tgss & Double & mJy beam$^{-1}$ &
Peak flux density at 150 MHz \\
E\_Peak\_flux\_tgss & Double & mJy beam$^{-1}$ &
Uncertainty in peak flux density \\

Maj\_tgss & Float & arcsec &
Major axis (FWHM) \\
E\_Maj\_tgss & Float & arcsec &
Uncertainty in major axis \\

Min\_tgss & Float & arcsec &
Minor axis (FWHM) \\
E\_Min\_tgss & Float & arcsec &
Uncertainty in minor axis \\

PA\_tgss & Float & deg &
Position angle east of north \\
E\_PA\_tgss & Float & deg &
Uncertainty in position angle \\

RMS\_noise & Float & mJy beam$^{-1}$ &
Local rms noise \\

Source\_code & String & \dots &
Source structure classification \\

Mosaic\_name & String & \dots &
TGSS mosaic image identifier \\
\hline
\multicolumn{4}{c}{Columns for VIKING \citep{edge2013}} \\
Z/Y/J/H/KsPETROMAG & Float & mag & total Z/Y/J/H/Ks-band flux\\
Z/Y/J/H/KsPETROMAGERR & Float & mag & total Z/Y/J/H/Ks-band flux error\\
\hline
\multicolumn{4}{c}{Columns for WISE \citep{wright2019allwise}}\\
w1/2/3/4mpro & Float & mag & total W1/2/3/4-band flux\\
w1/2/3/4sigmpro & Float & mag & total W1/2/3/4-band flux error\\
\hline
\multicolumn{4}{c}{Columns calculated in this work} \\
HSC\_dist  & Double & arcsec & The separation between VLASS source and HSC counterpart \\
on\_edge & Boolean & & 'True' if the source has a HSC--VLASS angular separation $>1''$\\
%has\_lotss & Boolean &  & The source is within the footprint of LoTSS \\
has\_lotss\_flux & Boolean & \dots & Source within the LoTSS catalog \\
has\_tgss & Boolean &  & The source is within the footprint of TGSS ADR1 \\
has\_tgss\_flux & Boolean & \dots & Source within the TGSS ADR1 catalog \\
has\_viking & Boolean & & The source is within the footprint of VIKING \\
has\_ukidss & Boolean & & The source is within the footprint of UKIDSS LAS\\
%sdss\_spec\_z & Double & & Redshift data from SDSS DR19. NaN = Not covered within the SDSS catalog.\\
redshift & Double & & Redshift data \\
redshift\_source & String & & Spectroscopic redshift, photometric redshift or dropout based approximation\\
alpha & Double & & Spectral index calculated \\
alpha\_source & String & & Source of calculating the spectral index\\
dropout & String & & The dropout group to which the source belongs \\

\hline

\end{longtable}

%\section{AAA}\label{sec:appendix}

%% I am leaving several commented out texts here.




%\vspace{8mm}
%\clearpage
\bibliography{VLASSHSC}
\bibliographystyle{aasjournal}
%\end{thebibliography}



\end{document}